\documentclass[prb,12pt,tightenlines,showpacs]{revtex4}
\usepackage{graphicx}
\pdfoutput=1

\begin{document}

\title{Effective pair potentials for spherical nanoparticles}

\author{Ramses van Zon}

\affiliation{Chemical Physics Theory Group, Department of Chemistry,
University of Toronto, 80 Saint George Street, Toronto, Ontario,
Canada M5S 3H6}

\date{22 September 2008}

\begin{abstract}
An effective description for spherical nanoparticles in a fluid of
point particles is presented. The points inside the nanoparticles and
the point particles are assumed to interact via spherically symmetric
additive pair potentials, while the distribution of points inside the
nanoparticles is taken to be spherically symmetric and smooth.  The
resulting effective pair interactions between a nanoparticle and a
point particle, as well as between two nanoparticles, are then given
by spherically symmetric potentials.  If overlap between particles is
allowed, the effective potential generally has non-analytic points,
but for each effective potential the expressions for different
overlapping cases can be written in terms of one analytic auxiliary
potential.  Effective potentials for hollow nanoparticles (appropriate
e.g.\ for buckyballs) are also considered, and shown to be related to
those for solid nanoparticles. Finally, explicit expressions are given
for the effective potentials derived from basic pair potentials of
power law and exponential form, as well as from the commonly used
London-Van der Waals, Morse, Buckingham, and Lennard-Jones potential.
The applicability of the latter is demonstrated by comparison with an
atomic description of nanoparticles with an internal face centered
cubic structure.
\end{abstract}


\maketitle

\section{Introduction}
\label{sec:introduction}

Nanoparticles,\cite{SchwarzSafran00,Choietal02,Hwangetal04TangAdvani06}
quantum dots,\cite{Chenetal05} colloidal
suspensions,\cite{Verbergetal97,Glotzeretal04} and globular
proteins\cite{TenWoldeFrenkel97Pellicaneetal04} are examples of
physical systems in which small nanometer or micron-sized clusters of
particles are suspended in a fluid. Such systems have applications
ranging from material coatings to drug
delivery.\cite{Jurgonsetal06Arrueboetal06,capsids} For colloidal
systems, collective behavior has been the focus of much
research,\cite{BarratHansen,Glotzeretal04} while nanoclusters are
often studied as isolated objects,\cite{BhattacharjeeElimelech97,
RothBalasubramanya00,BalasubramanyaRoth01,BalettoFerrando05,
Szwackietal07Gopakumaretal07} despite interesting collective phenomena
such as the increased heat conductance in dilute nanoparticle
suspensions\cite{Choietal02} and self-assembly.\cite{Glotzeretal04}

To study the collective properties of nanoparticles in suspension, one
would expect that a detailed description of the internal structure of
the clusters is not necessary, especially if the nanoparticles are
more or less solid.  On the other hand, a description in terms of hard
spheres would probably be too crude for nanoparticles since typical
atomic interaction ranges are on the order of \AA ngstroms. The aim of
this paper is to give a general effective description of nanoparticles
which retains a level of detail beyond the hard sphere model and which
is intended to be used in the study of the collective behavior of
nanoparticles, either numerically or analytically.  The starting point
of the description is to assume that each nanoparticle is composed of
particles with fixed relative positions, interacting with the point
particles in the fluid and their counterparts in other nanoparticles
through spherically symmetric pair potentials.  It is furthermore
assumed that the nanoparticles may be modeled as spheres with a smooth
spherically symmetric density of constituents, which can be viewed as
a smoothing procedure for the interactions. In particular, solid and
hollow spheres of uniform density are considered in detail, since
these are suitable for describing solid nanoclusters and buckyballs
(or similar structures), respectively.  The spherical smoothing
procedure results in spherically symmetric effective interaction
potentials for nanoparticles and point particles, and consequently
leads to a description of a nanoparticle as a single particle instead
of as a collection of particles.

Similar approaches to the problem of constructing effective potentials
have been used before, but only for specific cases.\cite{Hamaker37,
Girifalco92, BhattacharjeeElimelech97,RothBalasubramanya00,
BalasubramanyaRoth01,SchwarzSafran00} The current paper is devoted to
the general method of deriving effective pair potentials for
nanoparticles from the basic pair potential of their constituents. The
possibility of overlapping and embedded particles is specifically
treated as well.

The paper is structured as follows.  In Sec.~\ref{sec:general}, the
general smoothing procedure is explained.  Properties of the resulting
effective potentials are explored in Sec.~\ref{sec:symmetry}, with
special consideration for the difference between non-overlapping and
overlapping particles, which results in a reformulation of the
non-analytic effective potentials in terms of analytic auxiliary
potentials.  In Sec.~\ref{sec:solid-hollow}, the formalism is extended
to include hollow nanoparticles.  For uniform solid and hollow
nanoparticle structures, explicit effective potentials for a
nanoparticle and a point particle and for different nanoparticles are
worked out in Sec.~\ref{sec:explicit} for the London-van der Waals
potential, the exponential potential, the Morse potential, the
(modified) Buckingham potential, and the Lennard-Jones potential.
Section \ref{sec:comparison} addresses the applicability of the
effective potentials by comparison with an atom-based nanoparticle
model. A discussion in Sec.~\ref{sec:discussion} concludes the paper.

\section{Smoothing procedure for nanoparticle potentials}
\label{sec:general}

Consider a classical system of point particles, representing a fluid,
and spherical clusters called nanoparticles. While in reality, a
nanoparticle is a cluster of a number of atoms, here each nanoparticle
will be modeled by a smooth internal density profile $\rho(x)$ that
depends on the distance $x$ from the center of the nanoparticle only
and which is strictly zero for $x>s$, where $s$ is the radius of the
spherical nanoparticle.  This approximation is motivated by the idea
that for spherical nanoparticles, the inhomogeneities due to the
discreteness of the atoms inside the nanoparticles should only have a
small influence on the effective nanoparticle potentials.  Given a
density profile $\rho(x)$, one can make contact with the picture of a
nanoparticle as a cluster of distinct atoms by interpreting $M =
\int_{\mathcal B_{s}}\!{\rm d}\mathbf x\:\rho(x)$ as the total number
of atoms inside the nanoparticle, where $x=|\mathbf x|$, and
$\mathcal{B}_s$ denotes that the integration over $\mathbf x$ is over
the volume of a ball of radius $s$ around zero.

\begin{figure}[t]
\includegraphics[width=.9\columnwidth]{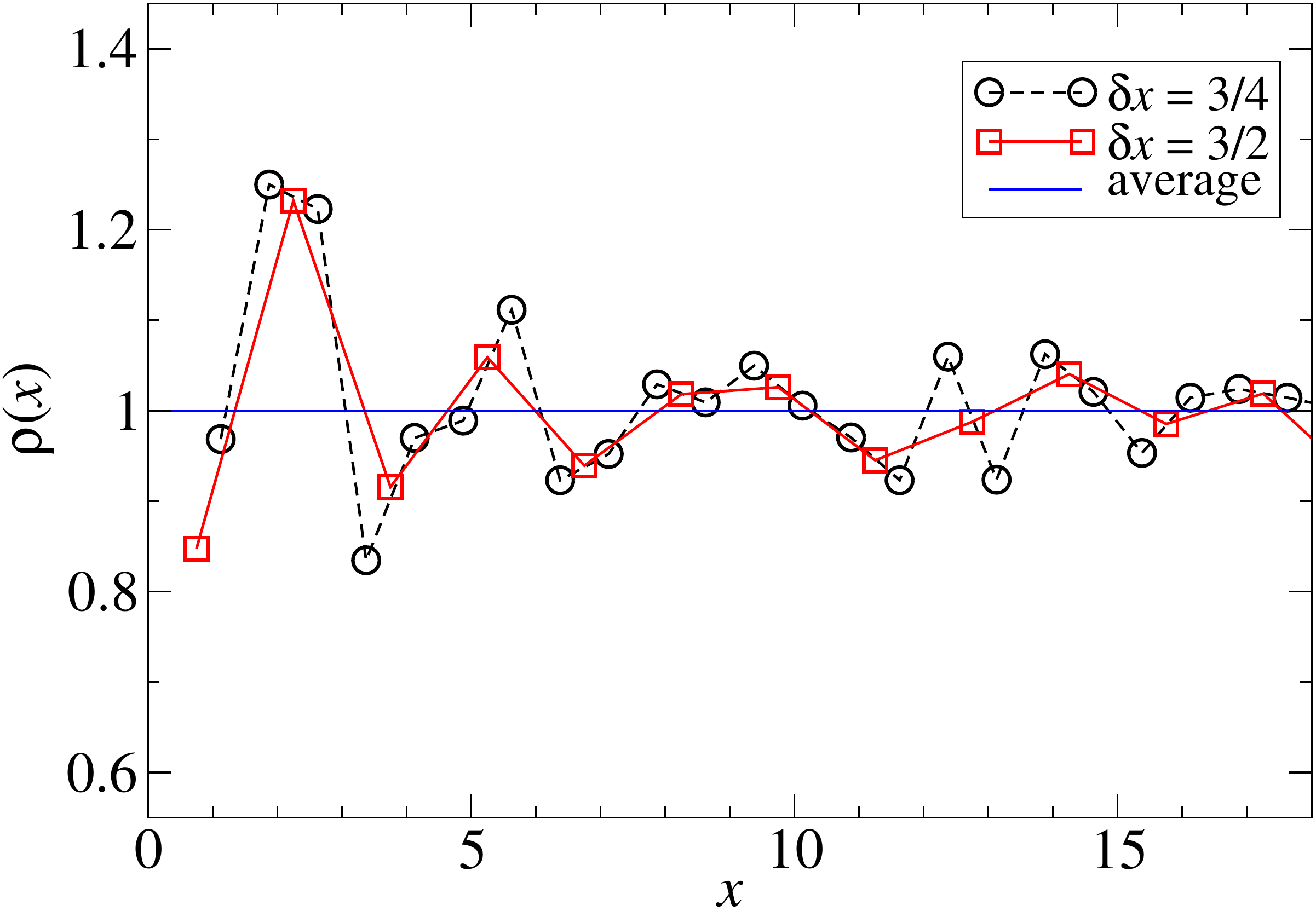}
\caption{Coarse-grained radial density profile of the fcc lattice of
  mean density $\bar\rho=1$ as a function of the distance from a
  central atom. The circles correspond to a coarse-graining width of
  $\delta x=3/4$, the squares corresponds to $\delta x=3/2$ (the
  points are connected to guide the eye).  The horizontal line
  indicates the mean number density. \label{smooth}}
\end{figure}

To further illustrate that it is reasonable to smooth out the internal
density, consider the idealized case that the atoms composing the
nanoparticle are arranged in a face-centered-cubic (fcc) lattice---the
crystal structure of e.g.\ aluminium, silver, gold, and
platinum\cite{Kittel86}---with one of the atoms in the center.  The
true density inside the nanoparticle is then a sum of delta functions,
but this can be coarse-grained by taking a spherical shell of radius
$x$ with a width $\delta x$, counting the number of atoms in the
shell, and dividing by the volume of the shell. The result of such
coarse-graining is shown in Fig.~\ref{smooth} for a lattice with mean
number density $\bar\rho=1$ and for two values of the coarse-graining
width, $\delta x=3/4$ and $\delta x=3/2$.  The coarse-grained density
around a single atom in an fcc crystal is seen to be reasonably
constant except near the central atom (with the positive and negative
deviations from the mean density averaging out for larger $\delta x$),
so that to first order the density may be replaced by a constant. This
highly idealized nanoparticle structure will be used again in
Sec.~\ref{sec:comparison} to get an idea of the accuracy of the
effective potentials.

Let $\phi_\mathrm{pn}(r)$ denote the basic pair potential between a
point of a nanoparticle and a point particle in the fluid, where $r$
is the distance between them. This potential will be assumed to be
analytic for $r>0$ but may diverge as $r\to0$. The effective
point-nanoparticle pair potential $V_\mathrm{pn}$ is then given by
\begin{equation}
    V_\mathrm{pn}(r)
    =  \int_{\mathcal{B}_s}\!\!{\rm d}\mathbf x\:
        \rho(x) \: \phi_\mathrm{pn}(|\mathbf r-\mathbf x|),   
\label{Ucgeneral}
\end{equation}
where the subscript pn denotes that this is a point
particle-nanoparticle potential and $\mathbf r$ is the distance vector
between the point particle and the center of the nanoparticle.
Because of the spherical symmetry of the density profile and the pair
potential $\phi_\mathrm{pn}$, the effective potential does not depend
on the direction of $\mathbf r$, only on its magnitude $r=|\mathbf
r|$.

Analogously, the effective inter-nanoparticle potential
$V_\mathrm{nn}$ for two nanoparticles with internal density profiles
$\rho_1$ and $\rho_2$, radii $s_1$ and $s_2$, and whose points
interact through a pair potential $\phi_\mathrm{nn}$, is given by
\begin{equation}
    V_\mathrm{nn}(r)
    =  \int_{\mathcal{B}_{s_1}}\!\!\! {\rm d}\mathbf x
        \int_{\mathcal{B}_{s_2}}\!\!\! {\rm d}\mathbf y\:
        \rho_1(x)\:\rho_2(y)\:
        \phi_\mathrm{nn}(|\mathbf r-\mathbf x-\mathbf y|),
\label{Uccgeneral}
\end{equation}
The potential $\phi_\mathrm{nn}$ will also be assumed to be analytic for
$r>0$.  Throughout this paper, $\phi_\mathrm{pn}$ and $\phi_\mathrm{nn}$
will be referred to as the \emph{basic pair potentials,} while
$V_\mathrm{pn}$ and $V_\mathrm{nn}$ are the \emph{effective potentials}.

To arrive at more concrete expressions for the effective potentials,
it will be assumed that the internal density profile of the
nanoparticles is analytic, so that it may be written as a Taylor
series,
\begin{equation}
    \rho(x) 
    = \Theta(s-x)\mathop{\sum_{i=0}}_{i\:\mathrm{even}}^\infty a_i x^i,
\label{nTaylor}
\end{equation}
where $\Theta$ is the Heaviside step function. In Eq.~(\ref{nTaylor}),
odd powers of $x$ were omitted since they lead to non-analytic
behavior at $x=0$.  The potentials for a nanoparticle and a point
particle, and for two nanoparticles, respectively, that would result
from internal densities of monomial form $\Theta(s-x) x^i$ are denoted
by
\begin{eqnarray}
    V_i(r) &=& \int_{B_s} \!\!{\rm d}\mathbf x\:
               x^i \: \phi_\mathrm{pn}(|\mathbf r+\mathbf x|),
\label{5}
\\
    V_{ij}(r) &=& \int_{\mathcal{B}_{s_1}}\!\! {\rm d}\mathbf x
                 \int_{\mathcal{B}_{s_2}}\!\! {\rm d}\mathbf y\:
       x^i\:y^j\:\phi_\mathrm{nn}(|\mathbf r+\mathbf x-\mathbf y|).
\label{6}
\end{eqnarray}
Here, and below, the dependence of $V_i$ and $V_{ij}$ on $s$ and $s_1$
and $s_2$ will not be denoted explicitly.  In terms of the potentials
$V_i$ and $V_{ij}$, the effective point-nanoparticle and
inter-nanoparticle potentials are given by
\begin{eqnarray}
    V_\mathrm{pn}(r)&=&\mathop{\sum_{i=0}}_{i\:\mathrm{even}}^\infty
    a_i V_i(r)
\label{Vnf}
\\
    V_\mathrm{nn}(r)&=&\mathop{\sum_{i=0}}_{i\:\mathrm{even}}^\infty 
                    \mathop{\sum_{j=0}}_{j\:\mathrm{even}}^\infty 
                    a_i b_j V_{ij}(r)
\label{Vnn}
\end{eqnarray}
where $\rho_1(x)=\Theta(s_1-x)\sum_i a_i x^i$ and $\rho_2(x) =
\Theta(s_2-x) \sum_j b_j x^j$ are the internal density profiles of two
interacting nanoparticles. While often only the first term $i=j=0$
will suffice, the formalism will be developed for general $i$ and $j$,
since this is not any more difficult.

The three-dimensional and six-dimensional integrals in Eqs.~(\ref{5})
and (\ref{6}) for the effective potentials make further manipulations
cumbersome.  However, due to the spherically symmetry of the basic
pair potentials, these multi-dimensional integrals can be rewritten as
integrals over a single variable.

To convert Eq.~(\ref{5}) to a single integral, one goes over to
spherical coordinates $\mathbf x=(x\sin\theta\cos\varphi,
x\sin\theta\sin\varphi, x\cos\theta)$, integrates over $\varphi$ and
then performs a change of integration variable from $\theta$ to $y =
[x^2\sin^2\theta + (r-x\cos\theta)^2]^{1/2}$, which yields
\begin{eqnarray*}
     V_i(r) 
     & = & \frac{2\pi}{r} \int_0^s\! {\rm d}x 
         \int_{|r-x|}^{r+x}\! {\rm d}y\: x^{i+1}\,y\,\phi_\mathrm{pn}(y) ,
\end{eqnarray*}
Reversing the order of the $x$ and $y$ integrals and using that $i$ is
even leads to
\begin{eqnarray}
    V_i(r)
    &=&\frac{2\pi}{(i+2)r}\,
      \Bigg[
      \int_{|r-s|}^{r+s}{\rm d}y\,[s^{i+2}-(r-y)^{i+2}]\,y\,\phi_\mathrm{pn}(y)
\nonumber\\&&+
      \Theta(s-r)\int_{0}^{s-r}\!\!{\rm d}y[(r+y)^{i+2}-(r-y)^{i+2}]
                                          y\phi_\mathrm{pn}(y)
      \Bigg]
.\nonumber\\
\label{this}
\end{eqnarray}
Defining a kernel
\begin{equation}
    K_i(x,s) = \frac{2\pi}{i+2}\, (s^{i+2}-x^{i+2})\,
    \Theta(s-|x|),
\label{kidef}
\end{equation}
one can write the right hand side of Eq.~(\ref{this}) in the concise
form
\begin{eqnarray}
    V_i(r)&=&\frac{1}{r}\int{\rm d}y\:K_i(r-y,s)\,y\,\phi_\mathrm{pn}(|y|),
\label{UsGeneral}
\end{eqnarray}
at least for $r>s$. That Eq.~(\ref{UsGeneral}) also holds for $r<s$
(with the same expression for $K_i$) is seen by writing the second
term in Eq.~(\ref{this}) as
\begin{eqnarray*}
&&
  \int_0^{s-r}\!\!{\rm d}y[ \{s^{i+2}-(r-y)^{i+2}\}
                         -\{s^{i+2}-(r+y)^{i+2}\}]y\phi_\mathrm{pn}(y)
\\
&&=
  \int_{-s+r}^{s-r}\!{\rm d}y\, [s^{i+2}-(r-y)^{i+2}]\,
                              y\,\phi_\mathrm{pn}(|y|).
\end{eqnarray*}
Combining this with the first term in Eq.~(\ref{this}) leads again to
Eq.~(\ref{UsGeneral}).  Note that for the special case of $i=0$, to be
used below, the kernel takes the form
\begin{equation}
    K_0(x,s) = \pi (s^2-x^2)\, \Theta(s-|x|).
\label{k0def}
\end{equation}

For the effective inter-nanoparticle potential $V_{ij}$, one can use
that the potential energy of two nanoparticles is equivalent to the
potential energy of a particle and a nanoparticle of which the points
interact via a point-nanoparticle potential $V_j$, i.e.,
\[
    V_{ij}(r)=\frac{1}{r}\int {\rm d}y\: K_i(r-y,s_1)\, y\, V_j(|y|),
\]
where in $V_j$, one should replace $s$ by $s_2$, and
$\phi_\mathrm{pn}$ by $\phi_\mathrm{nn}$.  Combining this with
Eq.~(\ref{UsGeneral}), and using that $K_j(x,s_2)$ is even in $x$, one
obtains
\begin{eqnarray}
    V_{ij}(r) &=& \frac{1}{r}\int\!{\rm d}y\,{\rm d}z\,
               K_i(r-y,s_1)\,K_j(y-z,s_2)\,z\,\phi_\mathrm{nn}(|z|),
\label{A}
\end{eqnarray}
or
\begin{eqnarray}
    V_{ij}(r)&=&\frac{1}{r}\int{\rm d}y\,K_{ij}(r-y,s_1,s_2)\,
                   y\,\phi_\mathrm{nn}(|y|),
\label{UssGeneral}
\end{eqnarray}
with the kernel $K_{ij}$ given by
\begin{equation}
    K_{ij}(x,s_1,s_2)
    = \int\! {\rm d}y \:K_i(x-y,s_1)\, K_j(y,s_2).
\label{Kijint0}
\end{equation}
The integral in this expression is further evaluated in the Appendix,
where it is shown that $K_{ij}$ is a piecewise polynomial function of
degree $i+j+5$ which has a finite support $|x|\leq s_1+s_2$, and
non-analytic points at $x=\pm |s_1-s_2|$. For the special case $i=j=0$
which will be used below, one finds from Eqs.~(\ref{Kij2}) and
(\ref{Kij3}), and after some rewriting,
\begin{equation}
    K_{00}(x,s_1,s_2)  = \left\{\begin{array}{ll}
                 \frac{\pi^2}{30}(D-|d |)^3(d ^2+3D|d |+D^2-5x^2) 
  	         &0\mbox{ if } |x|\leq|d|
               \\
	         \frac{\pi^2}{30}(D-|x|)^3(x^2+3D|x|+D^2-5d ^2)  
	         &\mbox{ if } |d |< |x| \leq D
               \\
	         0  
                 & \mbox{ if } |x| > D,
               \end{array}\right.
\label{K00def}
\end{equation}
where
\begin{eqnarray}
    D &=& s_1+s_2
\nonumber\\
    d &=& s_1-s_2. 
\label{Dddef}
\end{eqnarray}

Because the kernels $K_i$ and $K_{ij}$ are piecewise polynomials, the
integrals in Eqs.~(\ref{UsGeneral}) and (\ref{UssGeneral}) can be
performed analytically for many functional forms of $\phi_\mathrm{pn}$
and $\phi_\mathrm{nn}$, such as power law and exponential forms (see
Sec.~\ref{sec:explicit}), which are the basis of many commonly used
empirical pair potentials.

\section{Auxiliary potentials}
\label{sec:symmetry}

Although not evident from Eqs.~(\ref{UsGeneral}) and
(\ref{UssGeneral}), the non-analytic points of the kernels and of the
basic pair potential cause the effective potentials to have different
functional forms depending on whether there is overlap.  Different
overlapping cases can occur: A point particle and a nanoparticle can
either overlap (for $r<s$) or not overlap (for $r>s$), while two
nanoparticles can have no overlap, which requires $r>s_1+s_2=D$, or
partially overlap, or the smallest nanoparticle can be completely
embedded in the larger, which occurs when $r<|s_1-s_2|=|d|$.  The
different forms of the effective potentials for these different cases
can be linked by introducing auxiliary potentials.

The following symmetrization operations on functions $f$ are useful in
denoting the relations between effective and auxiliary
potentials:\cite{footnote}
\begin{eqnarray*}
  f([x]) &=& f(x) - f(-x) \mbox{ ``antisymmetrization''}
\\
  f((x)) &=& f(x) + f(-x) \mbox{ ``symmetrization.''}
\end{eqnarray*}
These operations are also useful for functions with multiple
arguments, e.g.,
\begin{eqnarray*}
  f([x], y)  &=& f(x, y) - f(-x, y)
\\
  f(x, (y))  &=& f(x, y) + f(x, -y)
\\
  f([x],[y]) &=& f(x, y) - f(-x, y) - f(x, -y) + f(-x, -y)
\\
  f([x, y])  &=& f(x, y) - f(-x, -y).
\end{eqnarray*}
Note that in the last example, a single antisymmetrization was
performed which involved both arguments.

The expressions of the effective potentials $V_i$ and $V_{ij}$ in
terms of auxiliary potentials (whose derivations will follow) are
given by
\begin{eqnarray}
    V_i(r) &=& 
    \left\{\begin{array}{ll}
      A_i((r), s) & \mbox{ if } r<s  \\
      A_i(r, [s]) & \mbox{ if } r>s
    \end{array}\right.
\label{auxiliaryi}
\\
    V_{ij}(r) &=&
    \left\{\begin{array}{ll}
      A_{ij}((r),[s_1],s_2) &\mbox{if $r<|d|$ and $s_1<s_2$}
      \\
      A_{ij}((r),s_1,[s_2]) &\mbox{if $r<d$ and $s_1>s_2$}
      \\
      A_{ij}((r),s_1,s_2)
                -A_{ij}(r,(s_1,-s_2)) &\mbox{if $|d|<r<D$}
      \\
      A_{ij}(r,[s_1],[s_2]) &\mbox{if $r>D$,}
    \end{array}\right.
\label{auxiliaryij}
\end{eqnarray}
in which the auxiliary potentials are defined as
\begin{eqnarray}
    A_i(r,s) 
    &=& \frac1r\int_0^{r+s}\!{\rm d}y\:
              \bar K_i(r-y,s)\, y\,\phi_\mathrm{pn}(y)
\label{Aidef}
\\
    A_{ij}(r,s_1,s_2)
    &=& \frac1r\int_0^{r+s_1+s_2}\!{\rm d}y\:
              \bar K_{ij}(r-y,s_1,s_2)\,
              y\, \phi_\mathrm{nn}(y),
\label{Aijdef}
\end{eqnarray}
where furthermore
\begin{eqnarray}
    \bar K_i(x,s) &=& \frac{2\pi}{i+2}\, (s^{i+2}-x^{i+2})
\label{barKidef}
\\
     \bar K_{ij}(x,s_1,s_2) &=&
     \int_{-s_2}^{x+s_1}\!{\rm d}y\:
     \bar K_i(x-y,s_1)\,\bar K_j(y,s_2).
\label{barKijdef0}
\end{eqnarray}
Note that $\bar K_i$ is the analytic continuation of $K_i$, while the
quantity $\bar K_{ij}(x,s_1,s_2)$ has the same functional
form as the kernel $K_{ij}$ for $x<0$, $d<|x|<D$ (as it coincides with
case \emph{4} in the appendix). In particular, for $i=j=0$, one has
from Eq.~(\ref{K00def})
\begin{eqnarray}
    \bar K_{00}(x,s_1,s_2)
    &=&
    \frac{\pi^2}{30}\,(s_1+s_2+x)^3\,
         (x^2-3s_1x-3s_2x-4s_1^2-4s_2^2+12s_1s_2).
\end{eqnarray}

The derivation of Eq.~(\ref{auxiliaryi}) goes as follows. Consider
first the non-overlapping case $r>s$. In that case, the absolute value
sign in the argument of $\phi_\mathrm{pn}$ may be dropped in
Eq.~(\ref{UsGeneral}), since $r>s$ and $r-y<s$ [cf.\
Eq.~(\ref{kidef})] imply that $y>0$. Thus, the effective
point-nanoparticle potential can be written as
\begin{eqnarray}
    V_i(r) 
&=&
\frac{1}{r}\int{\rm d}y\:K_i(r-y,s)\,y\,\phi_\mathrm{pn}(y)
\nonumber\\&
=&
\frac{1}{r}\int_{r-s}^{r+s}\!\!{\rm d}y\:\bar K_i(r-y,s)\,y\,\phi_\mathrm{pn}(y)
\nonumber\\
    &=& 
    \frac{1}{r}\int_{0}^{r+s}\!\!{\rm d}y\:\bar
    K_i(r-y,s)\,y\,\phi_\mathrm{pn}(y)
    +
    \frac{1}{r}\int_{r-s}^{0}\!\!{\rm d}y\:\bar
    K_i(r-y,s)\,y\,\phi_\mathrm{pn}(y)
\nonumber\\
    &=& A_i(r,s)-A_i(r,-s)
\nonumber\\
    &=& A_i(r,[s]),
\label{aux}
\end{eqnarray}
For the case $r<s$, the argument in the $\phi_\mathrm{pn}$ function in
Eq.~(\ref{UsGeneral}) needs to be $-y$ for $y<0$, giving
\begin{eqnarray}
    V_i(r)&=&
    \frac{1}{r}\int_{0}^{r+s}\!\!{\rm d}y\:\bar
    K_i(r-y,s)\,y\,\phi_\mathrm{pn}(y)
    +
    \frac{1}{r}\int_{r-s}^{0}\!\!{\rm d}y\:\bar
    K_i(r-y,s)\,y\,\phi_\mathrm{pn}(-y)
\nonumber\\
    &=&
    \frac{1}{r}\int_{0}^{r+s}\!\!{\rm d}y\:\bar
    K_i(r-y,s)\,y\,\phi_\mathrm{pn}(y)
    -
    \frac{1}{r}\int_0^{s-r}\!\!{\rm d}y\:\bar
    K_i(-r-y,s)\,y\,\phi_\mathrm{pn}(y),
\label{thisexpr}
\end{eqnarray}
where a change of integration variable from $y$ to $-y$ was carried
out in the second integral, and it was used that $\bar K_i(y,s)$ is
even in $y$.  The first term on the right hand side of
Eq.~(\ref{thisexpr}) is equal to $A_i(r,s)$ in Eq.~(\ref{Aidef}),
while the second term equals $A_i(-r,s)$, so that
\begin{eqnarray}
     V_i(r) &=& A_i(r,s)+A_i(-r,s)
     \equiv A_i((r),s).
\label{Uioverlap}
\end{eqnarray}
Thus, although the effective potentials between a point particle and a
nanoparticle have different forms for non-overlapping and overlapping
situations [Eqs.~(\ref{aux}) and (\ref{Uioverlap}), respectively],
both can be written in terms of the auxiliary potential $A_i$, and one
obtains Eq.~(\ref{auxiliaryi}).

A technical difficulty must be mentioned here, namely, that the
integral defining the auxiliary potential in Eq.~(\ref{Aidef}) may not
converge, even when the linear combinations in Eq.~(\ref{auxiliaryi})
do. In such cases, one should strictly write the auxiliary potential
as a sum of a regular and a diverging part by replacing the lower
limit of the integral in Eq.~(\ref{Aidef}) by $\delta>0$, and
expanding the result in $\delta$. In the absence of overlap,
Eq.~(\ref{auxiliaryi}) must yield a finite result, i.e., the diverging
parts (negative powers of $\delta$ and possibly logarithmic terms)
must cancel, hence in that case it suffices to work with the regular
part of the auxiliary potential.  On the other hand, in case of
overlap, it is possible that the divergent parts do not cancel in
Eq.~(\ref{Uioverlap}), resulting in infinite effective potentials.  An
independent criterion for whether an effective potential is infinite
in overlapping cases can be constructed as follows. For a single
particle inside a nanoparticle, the effective potential becomes
infinite only if the divergence of the basic pair potential
$\phi_\mathrm{pn}$ at the origin is too strong. In particular, if
$\phi(r)\propto r^{-k}$ for small $r$ then the point-nanoparticle
potential is infinite for $k\geq3$, as is seen by considering a small
sphere around the particle, giving an integral of the form
$\int_{r<\delta} {\rm d}\mathbf r\:\phi(r)\propto \int_0^\delta\!{\rm
d}r\:r^2r^{-k}\sim\frac{\delta^{3-k}}{3-k}$, which diverges for
$k\geq3$ in the limit $\delta\to0$.  This result extends to
inter-nanoparticle potentials, which are also infinite if there is
overlap and the potential $\phi_\mathrm{nn}$ diverges no slower than
$r^{-3}$, i.e., the $V_{ij}(r)$ are finite for $r<D$ provided
$\phi_\mathrm{nn}(r)$ diverges for small $r$ slower than $r^{-3}$.
Given this criterion, the divergent part of an auxiliary potential is
not needed to determine whether the corresponding effective potential
is infinite.  Since the divergent parts are needed neither in
overlapping nor in non-overlapping cases, below, only the regular
parts of auxiliary potentials will be given.

To derive Eq.~(\ref{auxiliaryij}) for the effective potentials between two
nanoparticles, one starts by rewriting Eq.~(\ref{A}) to
\begin{eqnarray}
    V_{ij}(r) &=& \frac{1}{r}\int_{-s_1}^{s_1} {\rm d}y
                           \int_{-s_2}^{s_2} {\rm d}x\: 
                           \bar K_i(y,s_1)\, \bar
                           K_j(x,s_2)\, 
                           (r-x-y)\, \phi_\mathrm{nn}(|r-x-y|).
\label{key}
\end{eqnarray}
In this formulation, the integration domain is a rectangle in the
$(x,y)$ plane and the integrand has a diagonal non-analytic line at
$x+y=r$. This line may or may not cross the domain, which is what
gives rise to non-analyticity and the difference between overlapping
and non-overlapping effective potentials.  

\begin{figure}[t]
\includegraphics[width=.3\columnwidth]{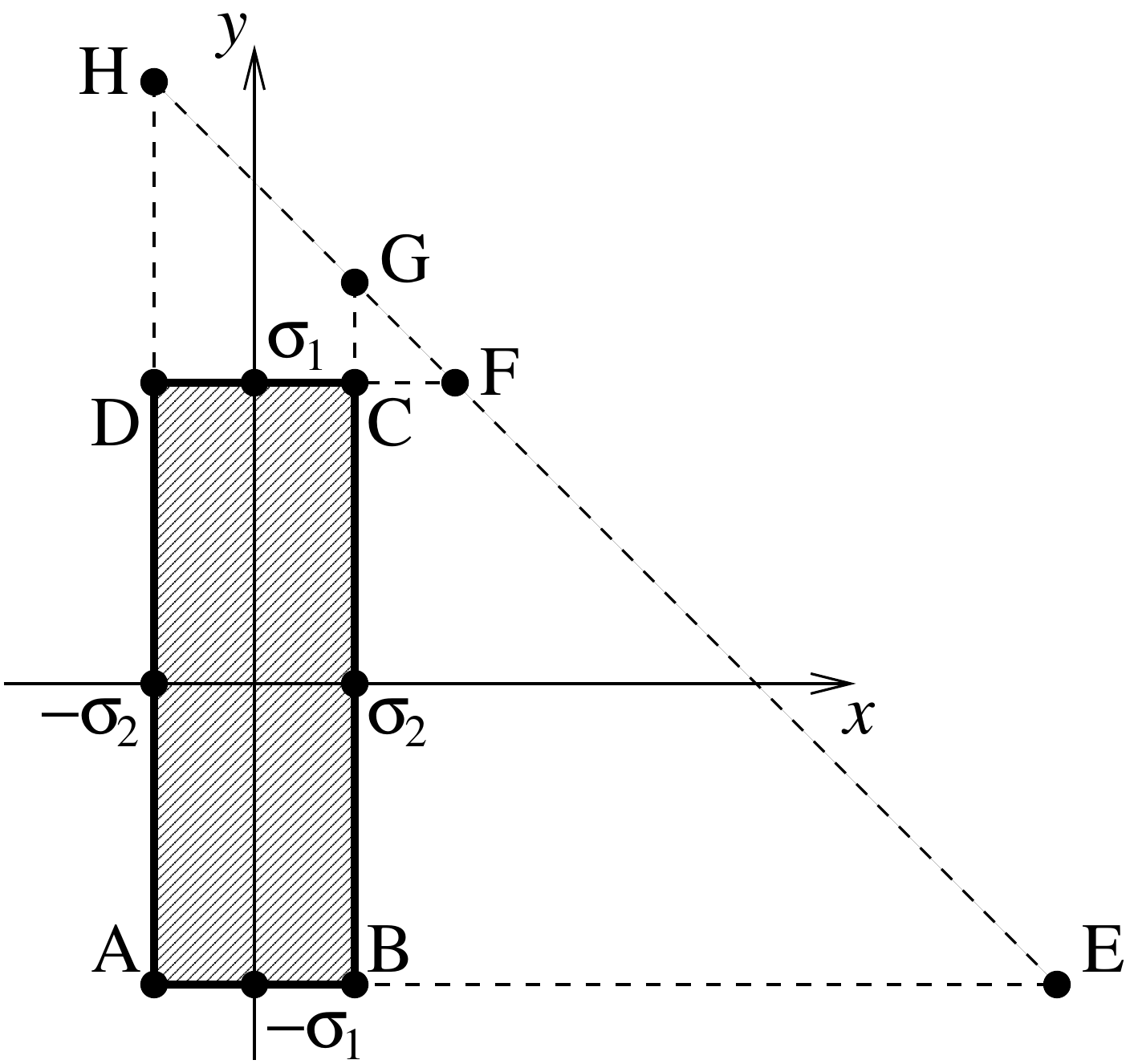}
\includegraphics[width=.3\columnwidth]{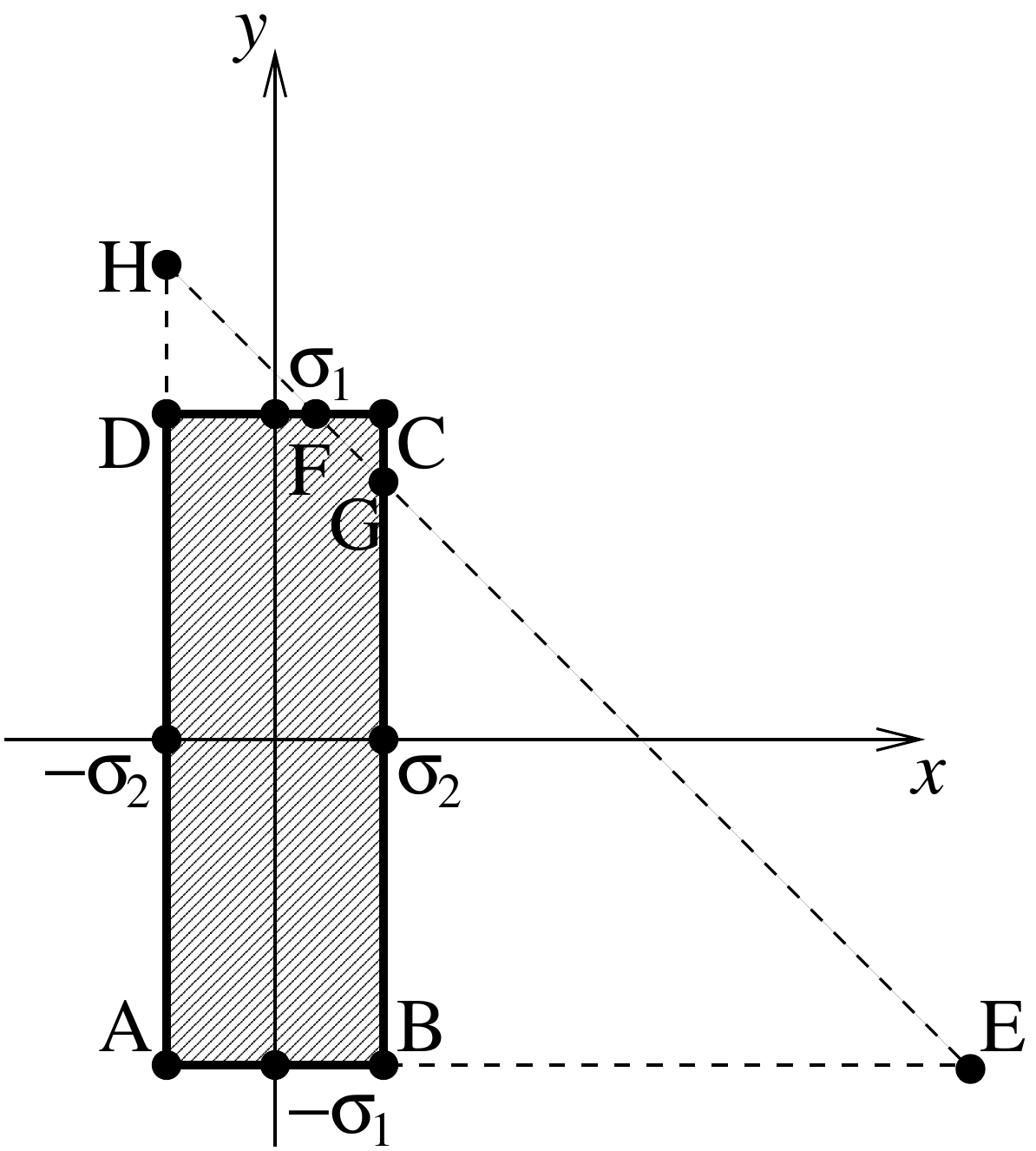}
\includegraphics[width=.3\columnwidth]{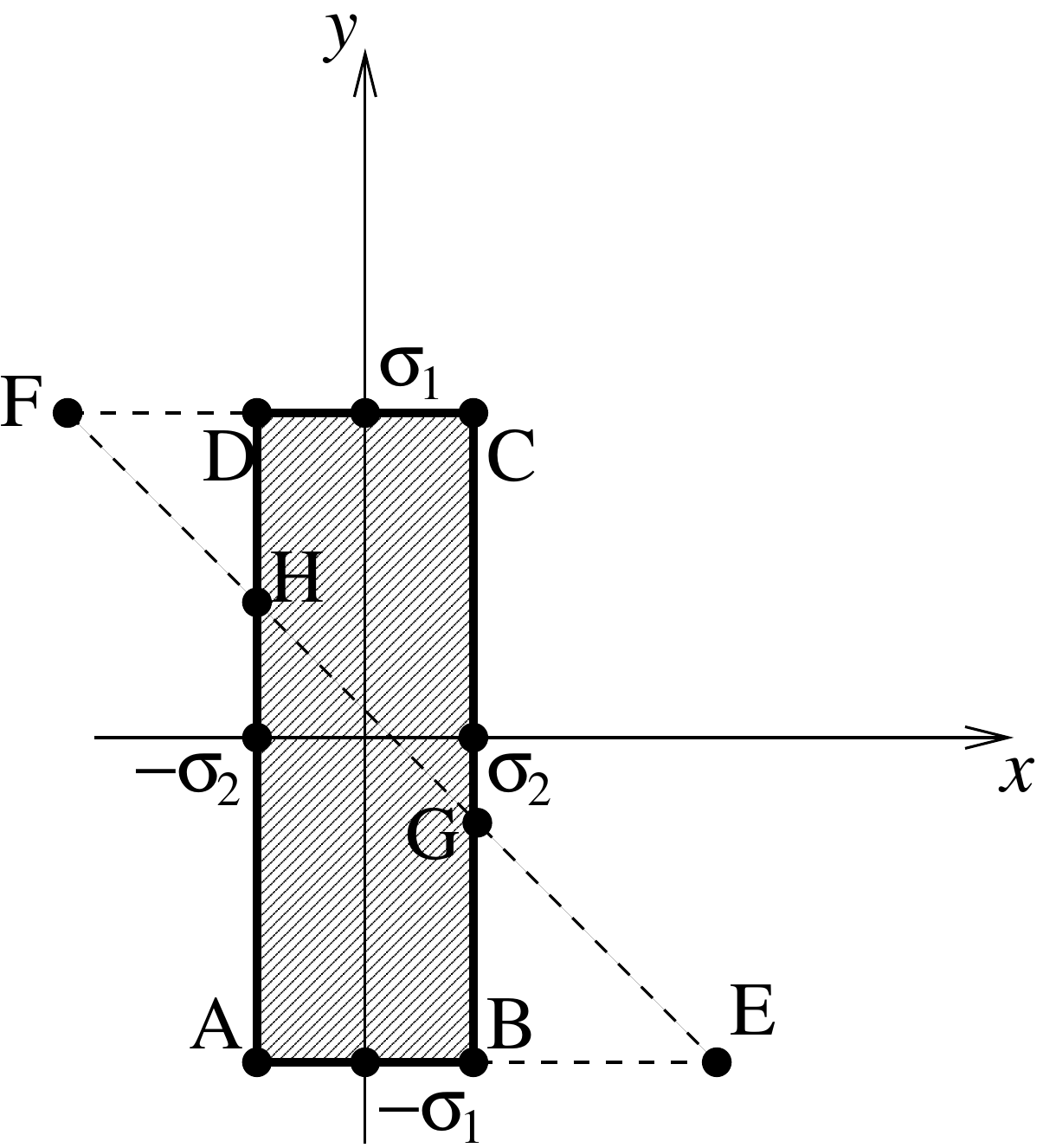}
\centerline{\hfill(a)\hfill(b)\hfill(c)\hfill}
\caption{Subdivision of the integration domain in the derivation of
  the expression of the inter-nanoparticle effective potential
  $V_{ij}$ in terms of the auxiliary potential $A_{ij}$. Assuming
  $s_1>s_2$, three cases have been distinguished: (a) $r>s_1+s_2$,
  (b) $s_1-s_2<r<s_1+s_2$, and (c) $r<s_1-s_2$.}
\label{fig:subdivisions}
\end{figure}

Subdividing the domain into triangular regions without
non-analyticities will result in expressions in terms of analytic
subexpressions.  The appropriate subdivisions of the integration
domain are shown in Fig.~\ref{fig:subdivisions}, where it was assumed
that the radius $s_1$ is larger than the radius $s_2$.  The
three panels of the figure correspond to the three cases that need to
be distinguished: (a) no overlap: $r>s_1+s_2$, (b) partial
overlap: $s_1-s_2<r<s_1+s_2$, and (c) complete overlap,
$r<s_1-s_2$.  In all three panels of
Fig.~\ref{fig:subdivisions}, the rectangle ABCD is the integration
domain, and the diagonal line through points E and H is the line of
non-analyticities (where $r-x-y=0$). For points below this line, the
absolute value in the argument of $\phi_\mathrm{nn}$ in Eq.~(\ref{key})
may be omitted, while for points above this line, it changes the sign
of the argument. Considering first case (a), i.e., no
overlap, one sees from Fig.~\ref{fig:subdivisions}(a) that
\begin{equation}
    V_{ij}(r) = I^+_{\rm AEH} - I^+_{\rm BEG} - I^+_{\rm DFH} +I^+_{\rm CFG},
\label{casea}
\end{equation}
where $I^+_{\rm XYZ}$ is the integral (\ref{key}) with the absolute value
sign omitted, and evaluated over the area of the triangle XYZ. For case
(b), i.e., partial overlap, one finds from
Fig.~\ref{fig:subdivisions}(b)
\begin{equation}
    V_{ij}(r) = I^+_{\rm AEH} - I^+_{\rm BEG} - I^+_{\rm DFH} +I^-_{\rm CFG},
\label{caseb}
\end{equation}
where the superscript ``$-$'' indicates that the sign of the argument of
$\phi_\mathrm{nn}$ in Eq.~(\ref{key}) is changed. Finally for case (c), one
finds from Fig.~\ref{fig:subdivisions}(c)
\begin{equation}
    V_{ij}(r) = I^+_{\rm AEH} - I^+_{\rm BEG} - I^-_{\rm DFH} +I^-_{\rm
    CFG}.
\label{casec}
\end{equation}
Note that for even basic potentials $\phi_\mathrm{pn}$ and
$\phi_\mathrm{nn}$, the sign of the arguments is inconsequential, so
that all three cases (\ref{casea})--(\ref{casec}) will have the same
functional form.

The integration limits appropriate for the triangular regions are
easily determined from Fig.~\ref{fig:subdivisions}. This yields the
following explicit expression for the integral $I^+_{\rm AEH}$:
\begin{eqnarray}
    I^+_{\rm AEH} 
    &=& \frac1r\int_{-s_2}^{r+s_1}\!{\rm d}x
             \int_{-s_1}^{r-x} \!{\rm d}y\:
             \bar K_i(y,s_1) \,\bar K_j(x,s_2)\, 
             (r-x-y)\, \phi_\mathrm{nn}(r-x-y)
\nonumber\\&&
             = A_{ij}(r,s_1,s_2).
\label{AijI}
\end{eqnarray}
Here, the identification with $A_{ij}$ followed from
Eqs.~(\ref{Aijdef}) and (\ref{barKijdef0}).  Given the form of the
auxiliary potential in Eq.~(\ref{AijI}), it is not hard to show that
\begin{eqnarray}
    I^+_{\rm BEG} = A_{ij}(r,s_1,-s_2),
\qquad
    I^+_{\rm DFH} = A_{ij}(r,-s_1,s_2),
\qquad
    I^+_{\rm CFG} = A_{ij}(r,-s_1,-s_2),
\end{eqnarray}
so that with Eq.~(\ref{casea}) one finds for the non-overlapping case 
\begin{eqnarray}
    V_{ij}(r) &=&
    A_{ij}(r,s_1,s_2)-A_{ij}(r,s_1,-s_2)
      -A_{ij}(r,-s_1,s_2)+A_{ij}(r,-s_1,-s_2)
\nonumber\\&
    =& A_{ij}(r,[s_1],[s_2]),
\label{toprove}
\end{eqnarray} 
As was the case for $A_i$, $A_{ij}$ may have divergent parts which
cancel in Eq.~(\ref{toprove}) and will be omitted below.

According to Eqs.~(\ref{casea}) and (\ref{caseb}), the partially
overlapping case (b) only requires replacing $I^+_{\rm CFG}$ by
$I^-_{\rm CFG}$, which is given by
\begin{eqnarray}
    I_{\rm CFG}^- &=& \frac1r\int_{r-s_1}^{s_2}\!\mathrm d x\int_{r-z}^{s_1}{\rm d}y\,
                           \bar K_i(y,s_1)\, \bar
                           K_j(x,s_2)\, 
                           (r-x-y)\, \phi_\mathrm{nn}(-r+x+y).
\end{eqnarray}
Substituting $y\to-y$, $x\to-x$, and using that $\bar K_i$ and $\bar K_j$
are even in $x$ and $y$, one finds
\begin{equation}
    I_{\rm CFG}^- = A_{ij}(-r,s_1,s_2),
\end{equation}
so that for $d<r<D$:
\begin{eqnarray}
    V_{ij}(r) &=&
    A_{ij}(r,s_1,s_2)-A_{ij}(r,s_1,-s_2)
                -A_{ij}(r,-s_1,s_2)+A_{ij}(-r,s_1,s_2)
\nonumber\\
    &= &
      A_{ij}((r),s_1,s_2)
                -A_{ij}(r,(s_1,-s_2))
\label{Uijpartial}
\end{eqnarray}
For the fully overlapping case, finally, one furthermore needs to
replace $I^+_{\rm DFH}$ by
\begin{eqnarray}
    I_{\rm DFH}^- 
    &=& \frac1r\int_{r-s_1}^{-s_2}\!\mathrm d x\int_{r-z}^{s_1}\!{\rm d}y\:
                           \bar K_i(y,s_1)\, \bar K_j(x,s_2)\, 
                           (r-x-y)\, \phi_\mathrm{nn}(-r+x+y)
\nonumber\\
    &=& A_{ij}(-r,s_1,-s_2),
\end{eqnarray}
whence for $r<d$:
\begin{eqnarray}
    V_{ij}(r) &=& A_{ij}(r,s_1,s_2)-A_{ij}(r,s_1,-s_2)
                -A_{ij}(-r,s_1,-s_2)+A_{ij}(-r,s_1,s_2)
\nonumber\\
    &=& A_{ij}((r),s_1,[s_2]).
\label{Uijtotal}
\end{eqnarray}
The reason that this is not symmetric in $s_1$ and $s_2$ is
because of the assumption that $s_1>s_2$. With
$s_1<s_2$ and $r<s_2-s_1$, one would have obtained
$V_{ij}(r) = A_{ij}((r),[s_1],s_2)$. This completes the
derivation of Eq.~(\ref{auxiliaryij}).

There is a degree of freedom in choosing the auxiliary potentials in
Eqs.~(\ref{auxiliaryi}) and (\ref{auxiliaryij}), since they enter
only in specific combinations.  In particular, according to
Eq.~(\ref{auxiliaryi}), the effective point-nanoparticle potential is
either $r$ symmetric or $s$-antisymmetric. Thus, one may replace
$A_i(r,s)$ by $A_i(r,s) + X(r,s)$ if the function $X(r,s)$ is
antisymmetric in $r$ as well as symmetric in $s$, i.e., if
\begin{equation}
X(r,s)  = X(r,-s) = - X(-r,s).
\label{freedomi}
\end{equation}
Conversely, any terms in $A_i$ that satisfy Eq.~(\ref{freedomi}) are
irrelevant to Eq.~(\ref{auxiliaryi}) and may, therefore, be omitted.
Similarly, the effective inter-nanoparticle potential in
Eq.~(\ref{auxiliaryij}) is not affected by adding a function
$Y(r,s_1,s_2)$ to the auxiliary potential $A_{ij}$, as long as $Y$
satisfies
\begin{eqnarray}
&&
  Y(r,s_1,s_2) - Y(r,-s_1,s_2) 
- Y(r,s_1,-s_2) + Y(r,-s_1,-s_2)  = 0
\nonumber\\&&
  Y(r,s_1,s_2) = Y(-r,-s_1,-s_2),
\label{freedomij}
\end{eqnarray}
while terms present in $A_{ij}$ that satisfy these relations are
irrelevant, and may be omitted.

\section{Solid and hollow nanoparticles}
\label{sec:solid-hollow}

Two particular cases of the internal nanoparticle densities $\rho$
will be considered in detail below.  The first is a uniform internal
density $\rho$ inside a solid sphere of radius $s$:
\begin{equation}
    \rho(x) = \rho\Theta(s-x).
\label{solid}
\end{equation}
Since Eq.~(\ref{solid}) is of the form $a_i\Theta(s-x) x^i$ with
$i=0$ and $a_0=\rho$, Eq.~(\ref{Vnf}) gives for the effective
point-nanoparticle potential
\begin{equation}
    V_\mathrm{pn}(r) = \rho V_0(r).
\end{equation}
Similarly, the effective inter-nanoparticle potential of two solid
nanoparticles of uniform density $\rho_1$ and $\rho_2$, and radii
$s_1$ and $s_2$, respectively, satisfies
[cf.~Eq.~(\ref{Vnn})]
\begin{equation}
    V_\mathrm{nn}(r)  = \rho_1 \rho_2 V_{00}(r).
\end{equation}

The second type of ``internal'' density $\rho(x)$ considered here is
that of hollow nanoparticles, whose density is concentrated on the
surface of the sphere, i.e.,
\begin{equation}
    \rho(x) = \tilde\rho\,\delta(s-x),   
\label{hollow}
\end{equation}
where $\tilde\rho$ is the surface density on the area of the sphere of
size $s$. This density is appropriate to describe e.g.\
buckyballs.\cite{Girifalco92} The density in Eq.~(\ref{hollow})
cannot be written in the form Eq.~(\ref{nTaylor}), but it is linked
to the uniform internal density in Eq.~(\ref{solid}) by
\begin{equation}
    \tilde\rho\delta(s-x)= \tilde\rho\,\frac{\partial\Theta(s-x)}{\partial s}.
\end{equation}
Consequently, the effective point-nanoparticle potential for this case
is given by
\begin{equation}
    V_\mathrm{pn}(r) = \tilde\rho\, V_\mathrm{h}(r),
\label{Vhdef0}
\end{equation}
with
\begin{equation}
    V_\mathrm{h}(r) = \frac{\partial V_0(r)}{\partial s},
\label{Vhdef}
\end{equation}
where the subscript h indicates that this potential acts between a
hollow nanoparticle and a point particle.

In a similar fashion, the inter-nanoparticle potentials for a solid
and a hollow nanoparticle (sh) is given by
\begin{equation}
    V_\mathrm{nn}(r) = \rho_1 \tilde\rho_2\, V_\mathrm{sh}(r) 
\label{Vijdef}
\end{equation}
and the potential for two hollow nanoparticles (hh) satisfies
\begin{equation}
    V_\mathrm{nn}(r) = \tilde\rho_1 \tilde\rho_2\, V_\mathrm{hh}(r) ,
\label{hh}
\end{equation}
where $\tilde\rho_1$ and $\tilde\rho_2$ are the surface density of the
two nanoparticles, while the scaled inter-nanoparticle potentials in
Eqs.~(\ref{Vijdef})--(\ref{hh}) are given by
\begin{eqnarray}
    V_\mathrm{sh}(r)&=&\frac{\partial V_{00}(r)}{\partial s_2}
\nonumber\\
\label{Urelations}
    V_\mathrm{hh}(r)&=&\frac{\partial^2V_{00}(r)}
                         {\partial s_1\partial s_2}.
\end{eqnarray}
Thus, the effective potentials $V_\mathrm{h}$, $V_\mathrm{sh}$ and
$V_\mathrm{hh}$ can be found by differentiation once $V_0$ and $V_{00}$,
are known.

The effective potentials for solid nanoparticles can be expressed in
terms of auxiliary potentials $A_0$ and $A_{00}$ using
Eqs.~(\ref{auxiliaryi}) and (\ref{auxiliaryij}).  In applying
Eqs.~(\ref{Vhdef}) and (\ref{Urelations}) to these expressions, it
should be realized that taking a derivative turns an antisymmetrized
function into a symmetrized one, and vice versa. Thus, by defining
\begin{eqnarray}
    A_\mathrm{h}(r,s)&=&\frac{\partial A_0(r,s)}{\partial s}
\nonumber
\\
    A_\mathrm{sh}(r,s_1,s_2)&=&\frac{\partial A_{00}
      (r,s_1,s_2)}{\partial s_2}
\label{Ahdef}
\\
    A_\mathrm{hh}(r,s_1,s_2)&=&\frac{\partial^2A_{00}
      (r,s_1,s_2)}
                         {\partial s_1\partial s_2},
\nonumber
\end{eqnarray}
one gets for the effective potentials
\begin{eqnarray}
    V_\mathrm{h}(r) &=& 
    \left\{\begin{array}{ll}
      A_\mathrm{h}((r),s)&\mbox{if $r<s$}
      \\
      A_\mathrm{h}(r,(s))&\mbox{if $r>s$,}
    \end{array}\right.
\label{auxiliaryh}
\\
    V_\mathrm{sh}(r) &= &
    \left\{\begin{array}{ll}
      A_\mathrm{sh}((r),[s_1],s_2) &\mbox{if $r<|d|$ and $s_1<s_2$}
\\
      A_\mathrm{sh}((r),s_1,(s_2)) &\mbox{if $r<d$ and $s_1>s_2$}
\\
      A_\mathrm{sh}((r),s_1,s_2)
      +A_\mathrm{sh}(r,[s_1,-s_2])
       &\mbox{if $|d|<r<D$}
\\
      A_\mathrm{sh}(r,[s_1],(s_2)) &\mbox{if $r>D$,}
    \end{array}\right.
\label{auxiliarysh}
\\
    V_\mathrm{hh}(r) &= &
    \left\{\begin{array}{ll}
      A_\mathrm{hh}((r),(s_1),s_2) &\mbox{if $r<|d|$ and $s_1<s_2$}
\\
      A_\mathrm{hh}((r),s_1,(s_2)) &\mbox{if $r<d$ and $s_1>s_2$}
\\
      A_\mathrm{hh}((r),s_1,s_2) +A_\mathrm{hh}(r,(s_1,-s_2))
      &\mbox{if $|d|<r<D$}
\\
      A_\mathrm{hh}(r,(s_1),(s_2))   &\mbox{if $r>D$.}
    \end{array}\right.
\label{auxiliaryhh}
\end{eqnarray}

\section{Effective potentials for uniformly solid and hollow nanoparticles}
\label{sec:explicit}

\subsection{Power laws}
\label{sec:powerlaws}

Pair potentials of power law form
\begin{equation}
    \phi^{n}(r) = \frac{1}{r^{n}},
\label{powerlaw}
\end{equation}
with $n$ integer, are basic building blocks of many atomic and
molecular pair potentials, such as the Coulomb potential ($n=1$) and
the Lennard-Jones potential (a linear combination of $n=6$ and
$n=12$). Note that here and below, a superscript on a potential
represents an index, not a power.

The effective potential $V_0^{n}$ for a point particle and a solid
nanoparticle of radius $s$ whose points interact with the
particle through $\phi_\mathrm{pn}=\phi^{n}$ is given  in
terms of the auxiliary potential by Eq.~(\ref{auxiliaryi}).  The
auxiliary potential follows from Eqs.~(\ref{Aidef}), giving, for
general $n$,
\begin{eqnarray}
    A_0^{n}(r,s) 
    &=& 
    \frac\pi r\int_0^{r+s}\!\!\mathrm dy\:
    \frac{s^2-(r-y)^2}{y^{n-1}}
    =
    \frac{2\pi[r+(n-3)s]}{(n-2)(n-3)(n-4)\:r\:(r+s)^{n-3}},
\label{auxk}
\end{eqnarray}
where divergent terms were omitted, as explained in
Sec.~\ref{sec:symmetry}.

The right hand side of Eq.~(\ref{auxk}) becomes ill-defined for the
specific values $n=2$, $3$ and~$4$.  This is caused by a term
proportional to $x^{n'-n-1}$ in the integrand in Eq.~(\ref{auxk})
(with $n'=2,$ 3 or 4), which when $n=n'$ should have resulted in a
term $\ln(r+s)$ instead of the erroneous and ill-defined expression
$\frac{(r+s)^{n'-n}}{n'-n}$ that occurs in Eq.~(\ref{auxk}).  Using
that $\lim_{n\to n'}\frac{\partial}{\partial n}[(n-n')
\frac{x^{n'-n}}{n'-n}] = \ln x$, this can be fixed by substituting
\begin{equation}
  A^{n'} \longrightarrow 
  \lim_{n\to n'} \frac{\partial}{\partial n}[(n-n')A^{n}].
\label{fixer}
\end{equation}
Applied to Eq.~(\ref{auxk}), this gives
\begin{eqnarray}
    A_0^{2}(r,s) &=& \frac{\pi(r+s)(3r-s)}{2r}+\frac{\pi(s^2-r^2)}{r}\ln(r+s)
\nonumber\\
    A_0^{3}(r,s) &=& -\frac{2\pi s}{r}+2\pi\ln(r+s)
\label{Ask234}\\
    A_0^{4}(r,s) &=& -\frac{\pi(3r+s)}{2r(r+s)}-\frac{\pi}{r}\ln(r+s).
\nonumber
\end{eqnarray}
The effective potential $V_0^{n}$ is obtained from these expressions
for the auxiliary potential using Eq.~(\ref{auxiliaryi}). 

{}From Eqs.~(\ref{Ahdef}) and (\ref{auxk}), it follows that the
auxiliary potential for a hollow nanoparticle and a point particle is
given by
\begin{equation}
    A_\mathrm{h}^{n}(r,s) = -\frac{2\pi s}{(n-2)\:r\:(r+s)^{n-2}}.
\label{Ahk}
\end{equation}
Equation (\ref{Ahk}) is ill-defined for $n=2$, in which case one uses
Eq.~(\ref{fixer}) to find
\begin{equation}
A_\mathrm{h}^{2}(r,s) = \frac{2\pi s}{r}\ln(r+s).
\label{Ahk2}
\end{equation}
The effective potential $V_h^{n}$ is now obtained from
Eq.~(\ref{auxiliaryh}).

For the effective inter-nanoparticle potential $V_{00}$, the auxiliary
potential formulation (\ref{auxiliaryij}) holds with $i=j=0$, where
the auxiliary potential is found using Eq.~(\ref{Aijdef}) with
$\phi_\mathrm{nn}=\phi^{n}$, giving
\begin{eqnarray}
    A_{00}^{n}(r,s_1,s_2) 
    &=&
    \frac{4\pi^2 p_n(r,s_1,s_2)}
         {(n-7)(n-6)(n-5)(n-4)(n-3)(n-2)\:r\:(r+s_1+s_2)^{n-5}}
,\label{Ik}
\end{eqnarray}
where
\begin{eqnarray}
p_n(r,s_1,s_2) &=& r^2+(n-5)(s_1+s_2)r
+(n-6)[s_1^2+s_2^2+(n-5)s_1s_2].
\end{eqnarray}
The expression in Eq.~(\ref{Ik}) is ill-defined for $n=2$, 3, 4, 5, 6
and 7. Using again Eq.~(\ref{fixer}), the correct expression for
$A_{00}^{n}$ for these values of $n$ is found to be
\begin{eqnarray}
    A_{00}^{n}(r,s_1,s_2)&=&
    \frac{4\pi^2}{r\:(r+s_1+s_2)^{n-5}\:
      \prod_{\stackrel{\mbox{\scriptsize$\ell=2$}}{\ell\neq n}}^7(\ell-n)}
\nonumber\\&&\times
    \bigg\{
      p_n(r,s_1,s_2)\Bigl[\ln(r+s_1+s_2)-
      \sum_{\stackrel{\mbox{\scriptsize$\ell=2$}}{\ell\neq
    n}}^7\frac{1}{\ell-n}
      \Big]
\nonumber\\&&\quad
          -s_1^2-s_2^2-(s_1+s_2)r+(11-2n)s_1s_2
    \bigg\}.
\label{Vss234567}
\end{eqnarray}

According to Eq.~(\ref{Ahdef}), the auxiliary potential for a solid
sphere of radius $s_1$ and a hollow sphere of radius $s_2$ can be
found by taking the derivative with respect to $s_2$, yielding, for
general $n$,
\begin{eqnarray}
    A_\mathrm{sh}^{n}(r,s_1,s_2) 
    &=&
    \frac{-4\pi^2s_2[r+(n-4)s_1+s_2]}{(n-5)(n-4)(n-3)(n-2)r(r+s_1+s_2)^{n-4}}
.\label{Ashk}
\end{eqnarray}
Finally, the effective potential for two hollow spheres follows from
another derivative with respect to $s_1$
[cf.~Eq.~(\ref{Ahdef})], leading to
\begin{eqnarray}
    A_\mathrm{hh}^{n}(r,s_1,s_2) &=& 
\frac{4\pi^2s_1s_2}{(n-3)(n-2)\:r\:(r+s_1+s_2)^{n-3}}
.\label{Ahhk}
\end{eqnarray}
For the ill-defined cases of Eqs.~(\ref{Ashk}) and (\ref{Ahhk}), one
can use Eq.~(\ref{fixer}) to get expressions similar to the one in
Eq.~(\ref{Vss234567}).

\subsection{Exponentials}
\label{sec:expcluster}

The effective interactions as a result of the exponential pair
potential
\begin{equation}
    \phi^\mathrm{E}(r) = e^{-r}
\label{exp}
\end{equation}
will now be derived.  Substituting this potential for $\phi_\mathrm{pn}$
in the expression (\ref{Aidef}) for the auxiliary potential gives
\begin{equation}
    A^\mathrm{E}_0(r,s) 
    = \frac{2\pi(3+r+s r+s^2+3s)}{r} e^{-r-s}
       +4\pi,
\label{AE0}
\end{equation}
where an irrelevant expression satisfying Eq.~(\ref{freedomi}) was
omitted.  {}From Eqs.~(\ref{Ahdef}) and (\ref{AE0}), the auxiliary
potential for a point particle and a hollow nanoparticle is found to
be
\begin{equation}
    A^\mathrm{E}_\mathrm{h}(r) = -\frac{2\pi s (1+r+s)}{r}e^{-r-s}.
\label{expA0h}
\end{equation}
Note that the corresponding effective potentials follow from
Eqs.~(\ref{auxiliaryi}) and (\ref{auxiliaryh}).

\begin{figure}[t]
\includegraphics[width=.48\columnwidth]{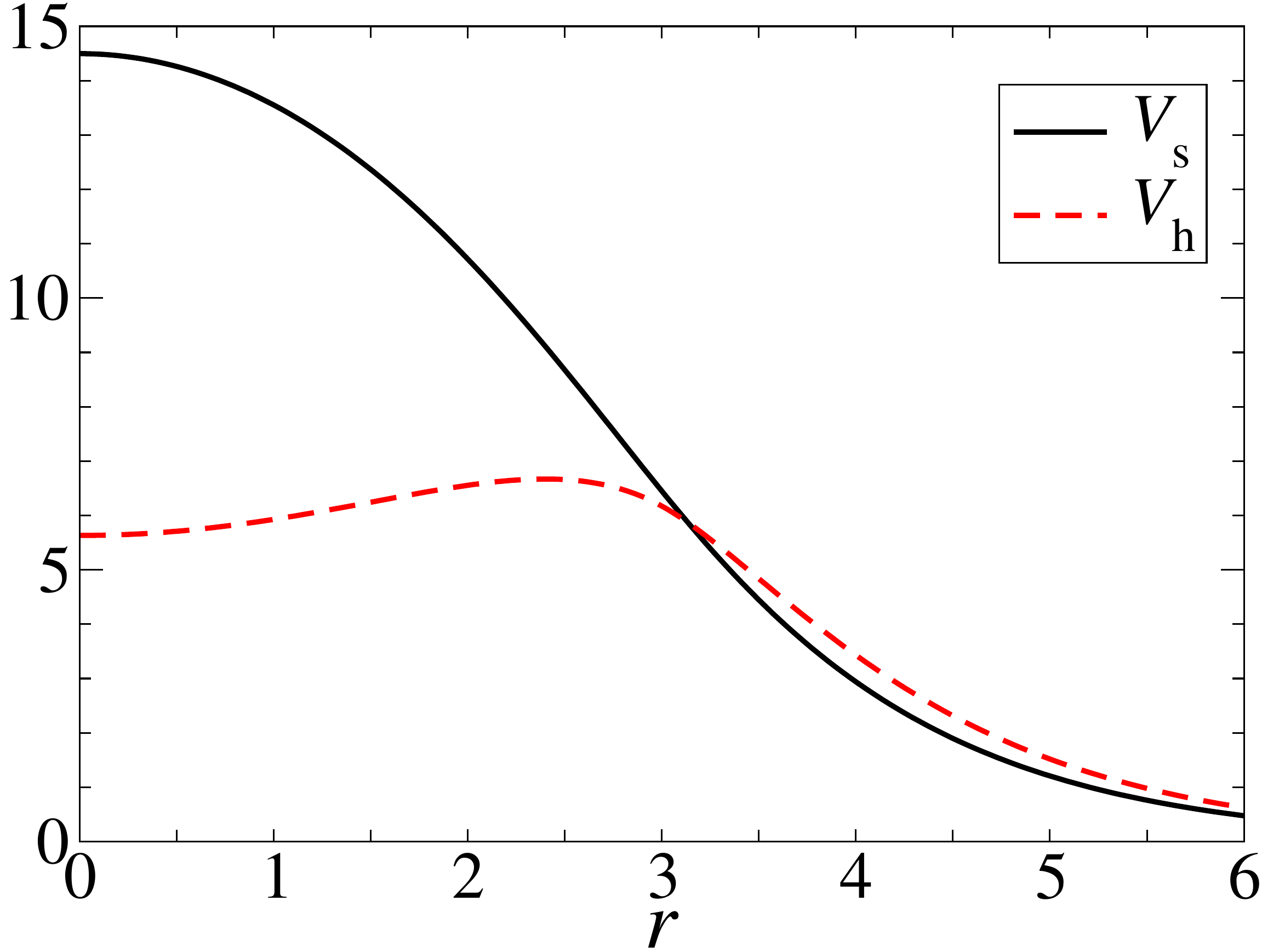}~ 
\includegraphics[width=.48\columnwidth]{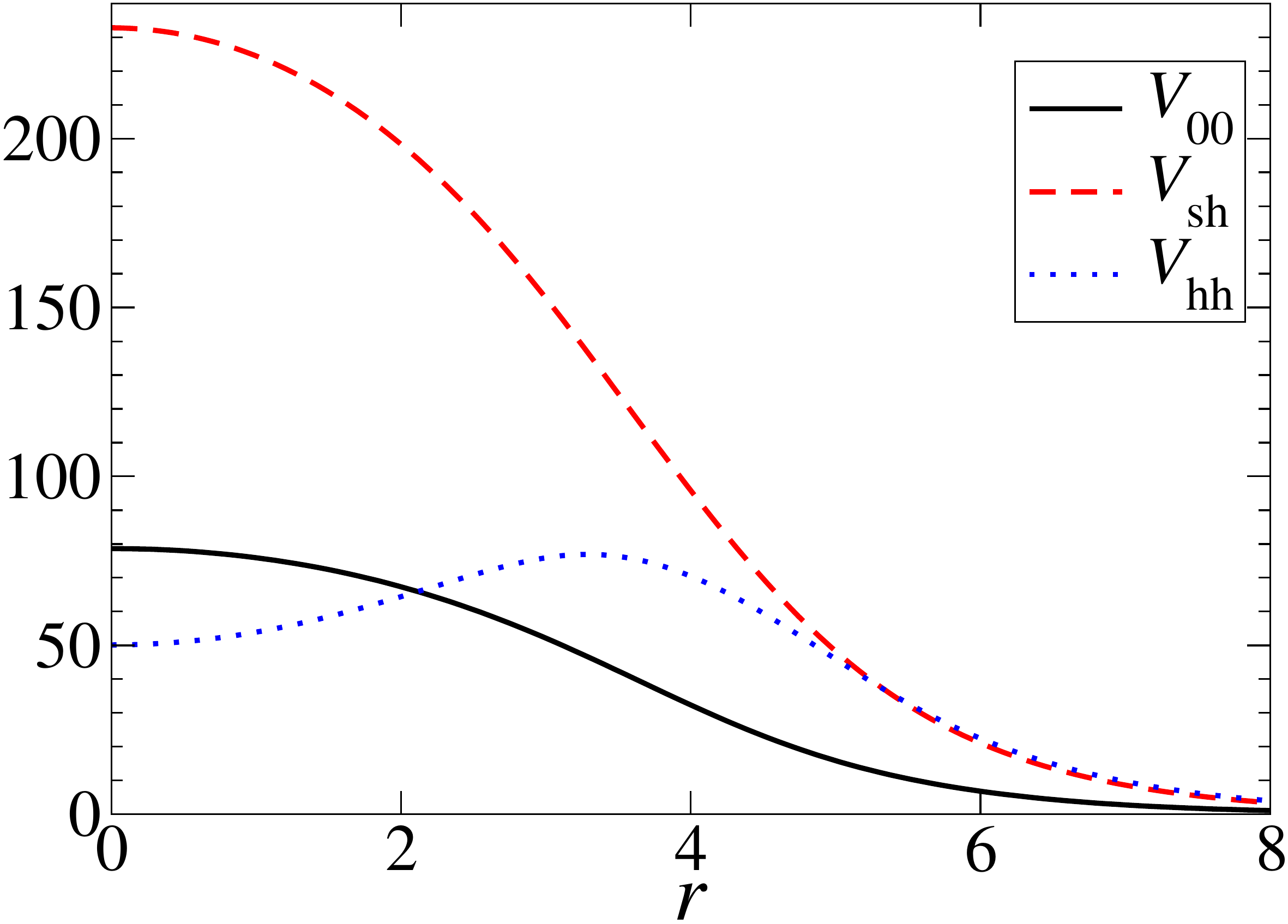}
\caption{Typical example of effective potentials based on an
exponential interaction [Eqs.~(\ref{auxiliaryij}) and (\ref{A00E})].
The left panel shows the point-nanoparticle potentials for solid (s)
and hollow (h) spheres with radius $s=3$, while the right panel
shows the inter-nanoparticle potentials for radii $s_1=4$ and
$s_2=1$.}
\label{Expplot}
\end{figure}

The effective inter-nanoparticle potential is of the auxiliary
potential form (\ref{auxiliaryij}) with $i=j=0$. The auxiliary
potential $A_{00}^\mathrm{E}$ is found using Eq.~(\ref{Aijdef}) with
$\phi_\mathrm{nn}=\phi^\mathrm{E}$, giving
\begin{widetext}
\begin{eqnarray}
&&
    A_{00}^\mathrm{E}(r,s_1,s_2) 
    =4\pi^2\frac{(r+s_1+s_2+5)(s_1+1)(s_2+1)+1-s_1s_2}
                {r}
                e^{-r-s_1-s_2}
\nonumber\\&&
   +\frac{\pi^2}{3r}\Bigg[
     8(s_1+s_2)(s_1^2+s_2^2-s_1s_2)r
     +6(s_1^2+s_2^2-4)(r^2+4)
     -r^4
     +3(s_1^2-s_2^2)^2+24
     \Bigg],
\label{A00E}
\end{eqnarray}
\end{widetext}
where an expression satisfying Eq.~(\ref{freedomij}) has been
omitted.  Using Eqs.~ (\ref{Ahdef}), the auxiliary potential for the
interaction between a solid and a hollow nanoparticle and between two
hollow particles are found to be
\begin{eqnarray}
    A_\mathrm{sh}^\mathrm{E}(r,s_1,s_2) 
    &=&\frac{-4\pi^2s_2[(r+s_1+s_2+4)(s_1+1)-s_1]}{r}e^{-r-s_1-s_2}
\nonumber\\*&&
   +\frac{4\pi^2s_2[(r+s_2)^2-s_1^2+4]}{r}
\label{AshE}
\\
    A_\mathrm{hh}^\mathrm{E}(r,s_1,s_2) 
   &=&\frac{4\pi^2s_1s_2(r+s_1+s_2+2)}{r}e^{-r-s_1-s_2}
   -\frac{8\pi^2s_1s_2}{r}.
\label{AhhE}
\end{eqnarray}

Figure \ref{Expplot} shows a typical example of the effective
potentials derived from the exponential basic potential
[cf.~Eqs.~(\ref{auxiliaryi}), (\ref{auxiliaryij}),
(\ref{auxiliaryh})--(\ref{auxiliaryhh}) and
(\ref{AE0})--(\ref{AhhE})].  One sees that these effective potentials
are very smooth and do not have a hard core, which is typical for
effective potentials based on a basic pair potential that does not
diverge for small distances.

\subsection{Examples using common pair potentials}
\label{common}

\begin{figure}[t]
\includegraphics[width=.48\columnwidth]{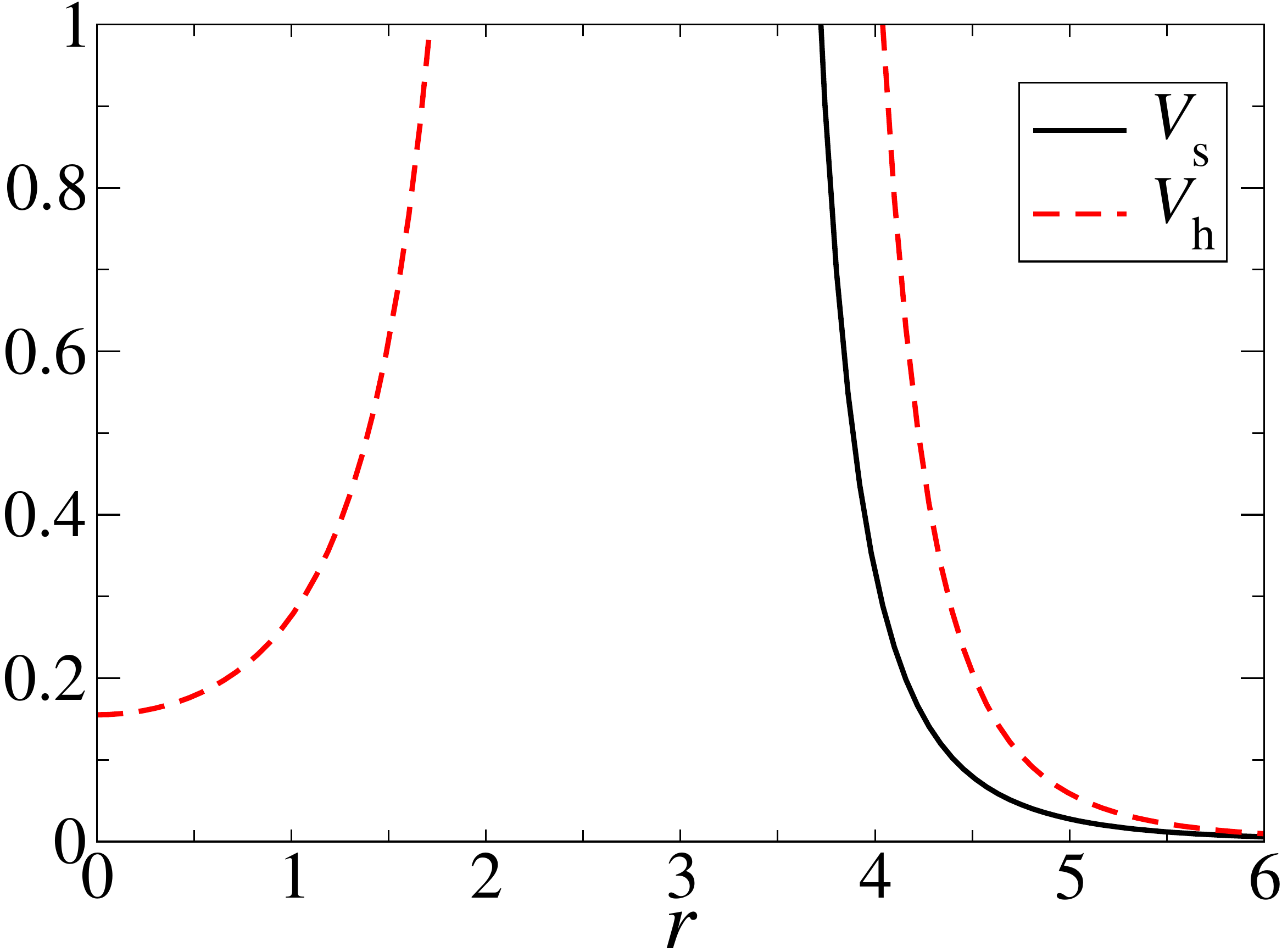}~ 
\includegraphics[width=.48\columnwidth]{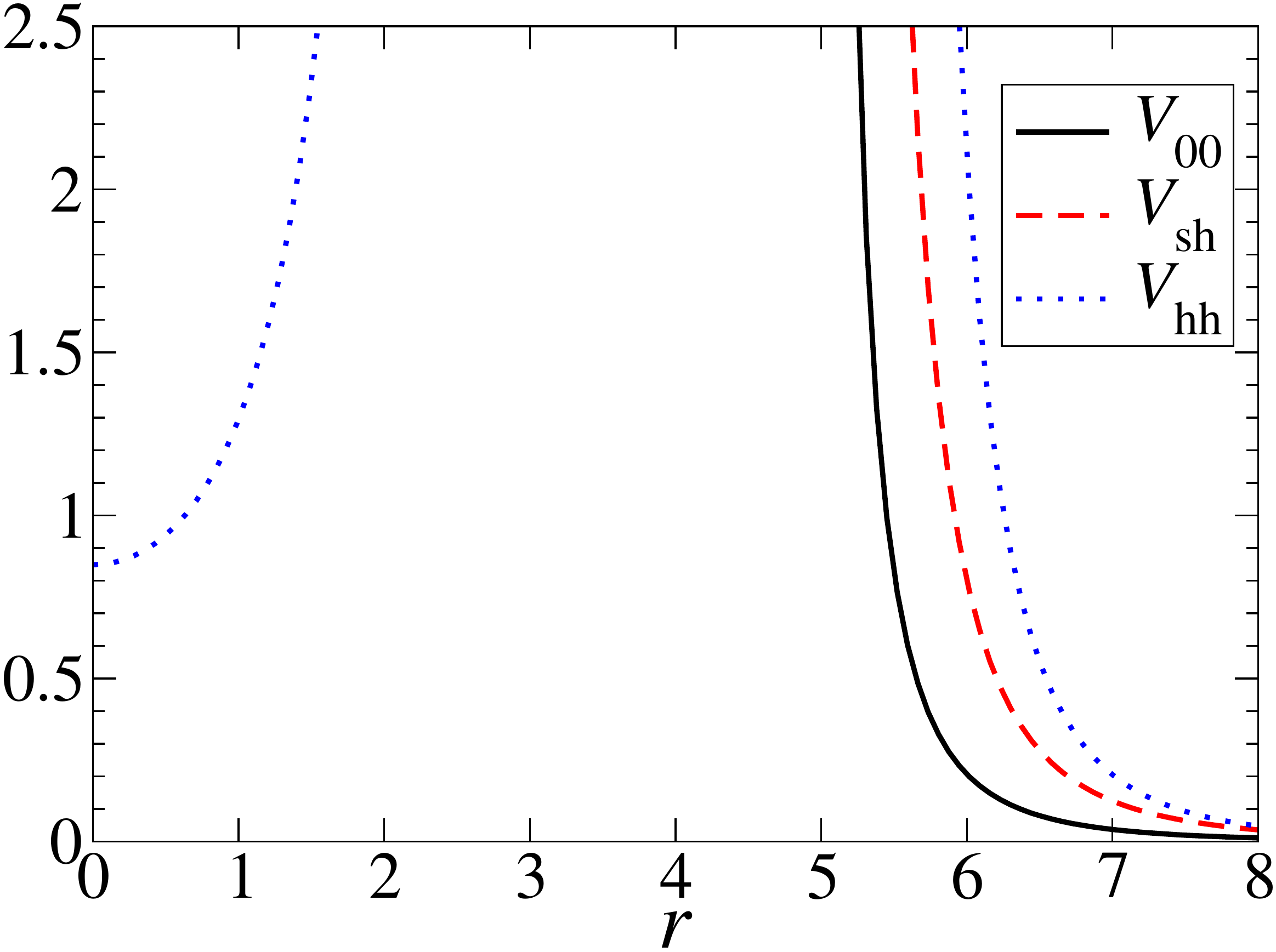}
\caption{Typical example of effective potentials based on the
  London-van der Waals interaction, i.e., the power law in
  Eq.~(\ref{powerlaw}) with $n=6$.  The left panel shows the
  potential for a point particle and solid or hollow nanoparticle of
  radius $s=3$, the right panel shows the potential for two
  nanoparticles of radius $s_1=4$ and $s_2=1$.}
\label{LvdWplot}
\end{figure}

\subsubsection*{London-van der Waals potential}

In this section, the effective potentials based on the London-van der
Waals potential
\begin{equation}
     \phi^{6}(r) = \frac{1}{r^{6}}
\label{LvdW}
\end{equation}
will be presented. Note that the negative prefactor that occurs in
front of the attractive London-van der Waals interaction has been
omitted here.  Substituting $n=6$ into Eq.~(\ref{auxk}), and using
Eq.~(\ref{auxiliaryi}), one finds the London-van der Waals potential
for a solid nanoparticle and a point particle:
\begin{equation}
     V^{6}_0(r) = \frac{4\pi s^3}{3(r^2-s^2)^3},
\label{vdwVs}
\end{equation}
for $r>s$.  This effective potential becomes infinite for
$r<s$.  For the London-van der Waals interaction of a hollow
nanoparticle with a point particle, Eqs.~(\ref{auxiliaryh}) and
(\ref{Ahk}) with $n=6$, lead to
\begin{equation}
    V^{6}_\mathrm{h}(r) = \frac{4\pi s^2}{(r^2-s^2)^3}
                    + \frac{8\pi s^4}{(r^2-s^2)^4}.
\label{vdwV02}
\end{equation}

The effective London-van der Waals interaction potential for two solid
nanoparticles is determined by substituting $n=6$ into
Eq.~(\ref{Vss234567}), and using Eq.~(\ref{auxiliaryij}), which gives
\begin{equation}
    V^{6}_{00}(r) = \frac{\pi^2s_1s_2}{3(r^2-d^2)}
                      +\frac{\pi^2s_1s_2}{3(r^2-D^2)}
                      +\frac{\pi^2}{6}\ln\frac{r^2-D^2}{r^2-d^2}.
\label{vdwV11}
\end{equation}
This result coincides with that of Hamaker.\cite{Hamaker37}

Using Eqs.~(\ref{Urelations}) and (\ref{vdwV11}), or using
Eqs.~(\ref{Ashk}) and (\ref{auxiliarysh}), one finds for the
London-van der Waals potential $V^{6}_\mathrm{sh}$ for a solid
nanoparticle of radius $s_1$ and a hollow nanoparticle of radius~$s_2$
\begin{eqnarray}
    V^{6}_\mathrm{sh}(r) &=& 
                       \frac{2\pi^2s_1s_2D}{3(r^2-D^2)^2}
                      -\frac{2\pi^2s_1s_2d}{3(r^2-d^2)^2}
                      -\frac{\pi^2s_2}{3(r^2-D^2)}
                      +\frac{\pi^2s_2}{3(r^2-d^2)}.
\label{vdwV12}
\end{eqnarray}

The effective London-van der Waals potential $V^{6}_\mathrm{hh}$ for two hollow
nanoparticles, finally, is obtained from
Eq.~(\ref{vdwV12}) using Eq.~(\ref{Urelations}), or alternatively
from Eqs.~(\ref{Ahhk}) and (\ref{auxiliaryhh}), with the result
\begin{eqnarray}
    V^{6}_\mathrm{hh}(r) &=& 
                       \frac{8\pi^2s_1s_2D^2}{3(r^2-D^2)^3}
                       -\frac{8\pi^2s_1s_2d^2}{3(r^2-d^2)^3}
                       +\frac{2\pi^2s_1s_2}{3(r^2-D^2)^2}
                       -\frac{2\pi^2s_1s_2}{3(r^2-d^2)^2}.
\label{vdwV22}
\end{eqnarray}
Figure \ref{LvdWplot} shows a typical example of the effective
potentials for the London-van der Waals interaction as the basic pair
potential.

\subsubsection*{Morse potential}

\begin{figure}[t]
\includegraphics[width=.48\columnwidth]{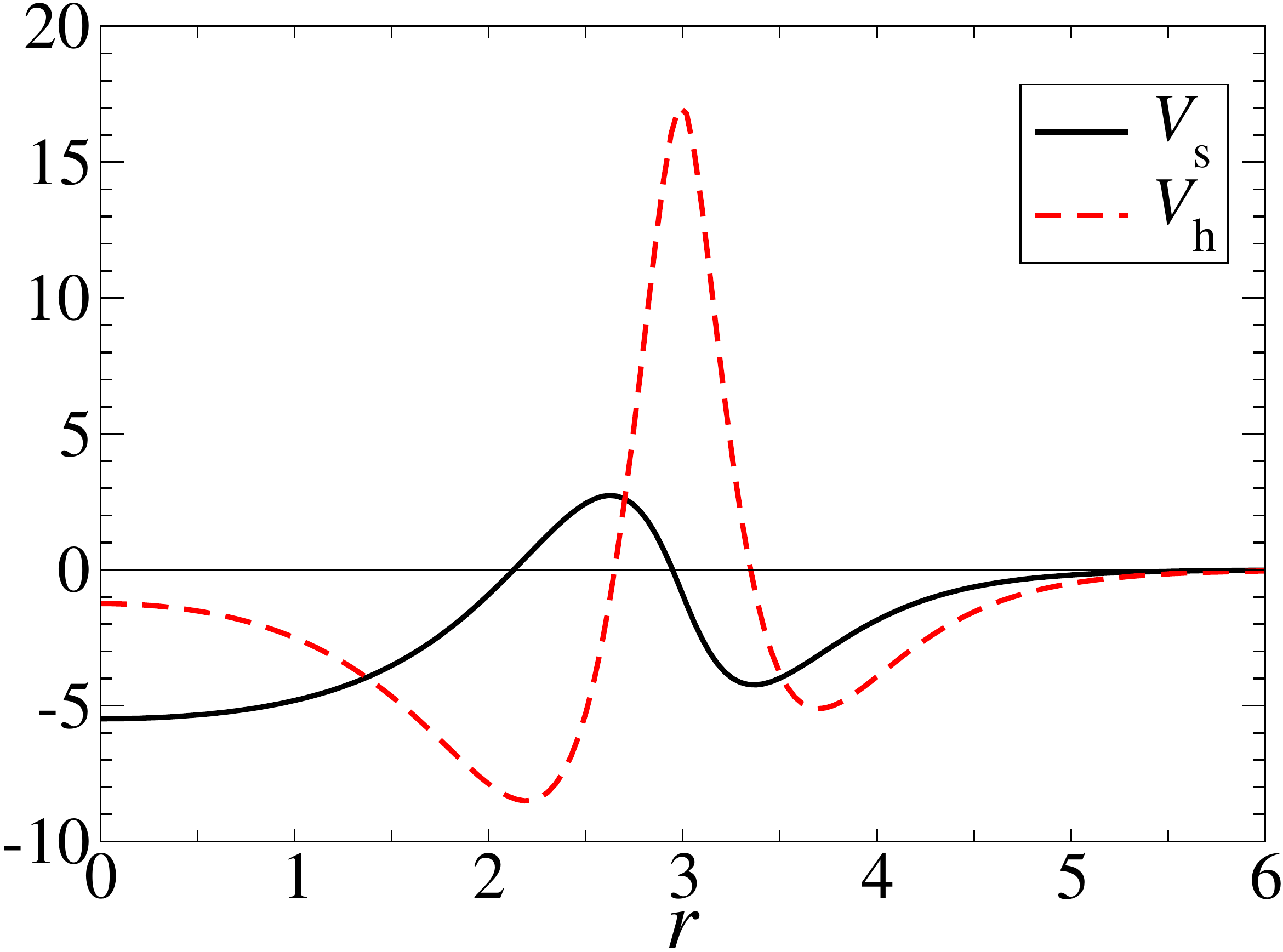}~ 
\includegraphics[width=.48\columnwidth]{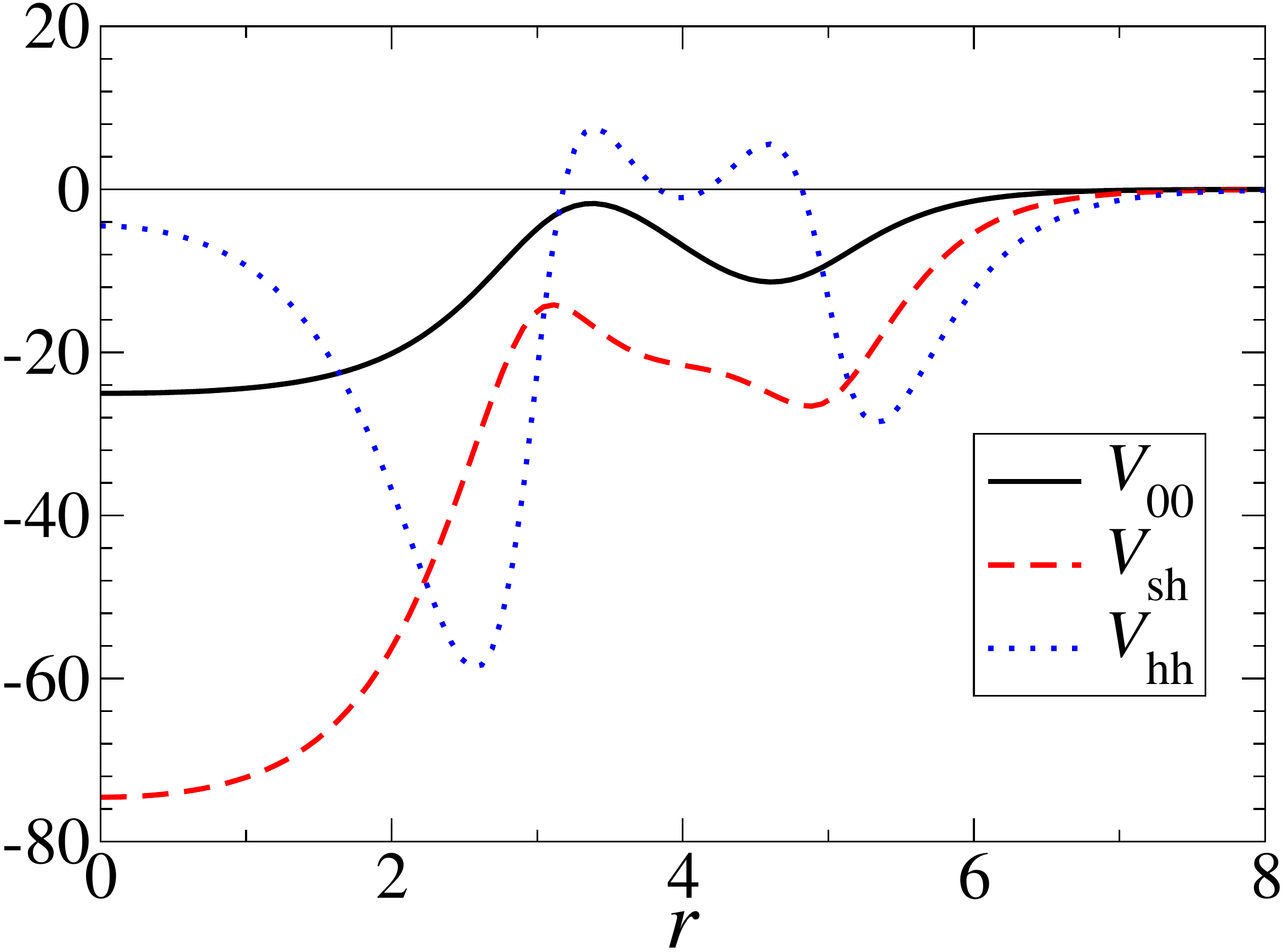}
\caption{Example of Morse effective potentials for $b=2.6$.  The left
  panel shows the effective potential for a particle and a solid or
  hollow nanoparticle of radius $s=3$, the right panel shows the
  effective potentials for two nanoparticles of radius $s_1=4$ and
  $s_2=1$.}
\label{Morseplot26}
\end{figure}

The Morse potential\cite{Morse29}
\begin{equation}
     \phi^\mathrm{M}(r) = e^{-2b(r-1)} -2 e^{-b(r-1)},
\label{Morse}
\end{equation}
is used e.g.\ for molecular bonds and for pure
metals.\cite{GirifalcoWeizer59} It is a sum of two exponential
functions, so having derived the formulas for the exponential
potential in Sec.~\ref{sec:expcluster}, one easily finds the
corresponding point-nanoparticle interactions by taking the
combinations
\begin{eqnarray}
    V_0^\mathrm{M}(r) 
    & =& \frac{e^{2b}}{2^3b^3}\,V_0^\mathrm{E}(2br,2bs) 
        - \frac{2e^{b}}{b^3}\,V_0^\mathrm{E}(br,bs)
\label{VMs}
\\
    V_\mathrm{h}^\mathrm{M}(r) 
    & = & \frac{e^{2b}}{2^2b^2}\,V_\mathrm{h}^\mathrm{E}(2br,2bs) 
       - \frac{2e^{b}}{b^2}\,V_\mathrm{h}^\mathrm{E}(br,bs),
\label{VMj}
\end{eqnarray}
where the notation $V_0^\mathrm{E}(\alpha r,\beta s)$ indicates
that in $V^\mathrm{E}_0$ and $V^\mathrm{E}_\mathrm{h}$, $r$ is to be
replaced by $\alpha r$ and $s$ by $\beta s$.  Likewise, the
inter-nanoparticle interactions for the Morse potential in
Eq.~(\ref{Morse}) are given by
\begin{eqnarray}
&&
  V_{00}^\mathrm{M}(r)
    = \frac{e^{2b}}{2^6b^6}\,V_{00}^\mathrm E(2br,2bs_1,2bs_2)
     - \frac{2e^{b}}{b^6}\,V_{00}^\mathrm{E}(br,bs_1,bs_2).
\label{VMss}
\\&&
    V_\mathrm{sh}^\mathrm{M}(r)
    = \frac{e^{2b}}{2^5b^5}\,V_\mathrm{sh}^\mathrm E(2br,2bs_1,2bs_2)
        - \frac{2e^{b}}{b^5}\,V_\mathrm{sh}^\mathrm{E}(br,bs_1,bs_2).
\label{VMsh}
\\&&
    V_\mathrm{hh}^\mathrm{M}(r)
    = \frac{e^{2b}}{2^4b^4}\,V_\mathrm{hh}^\mathrm E(2br,2bs_1,2bs_2)
       - \frac{2e^{b}}{b^4}\,V_\mathrm{hh}^\mathrm{E}(br,bs_1,bs_2).
\label{VMhh}
\end{eqnarray}

\begin{figure}[t]
\includegraphics[width=.48\columnwidth]{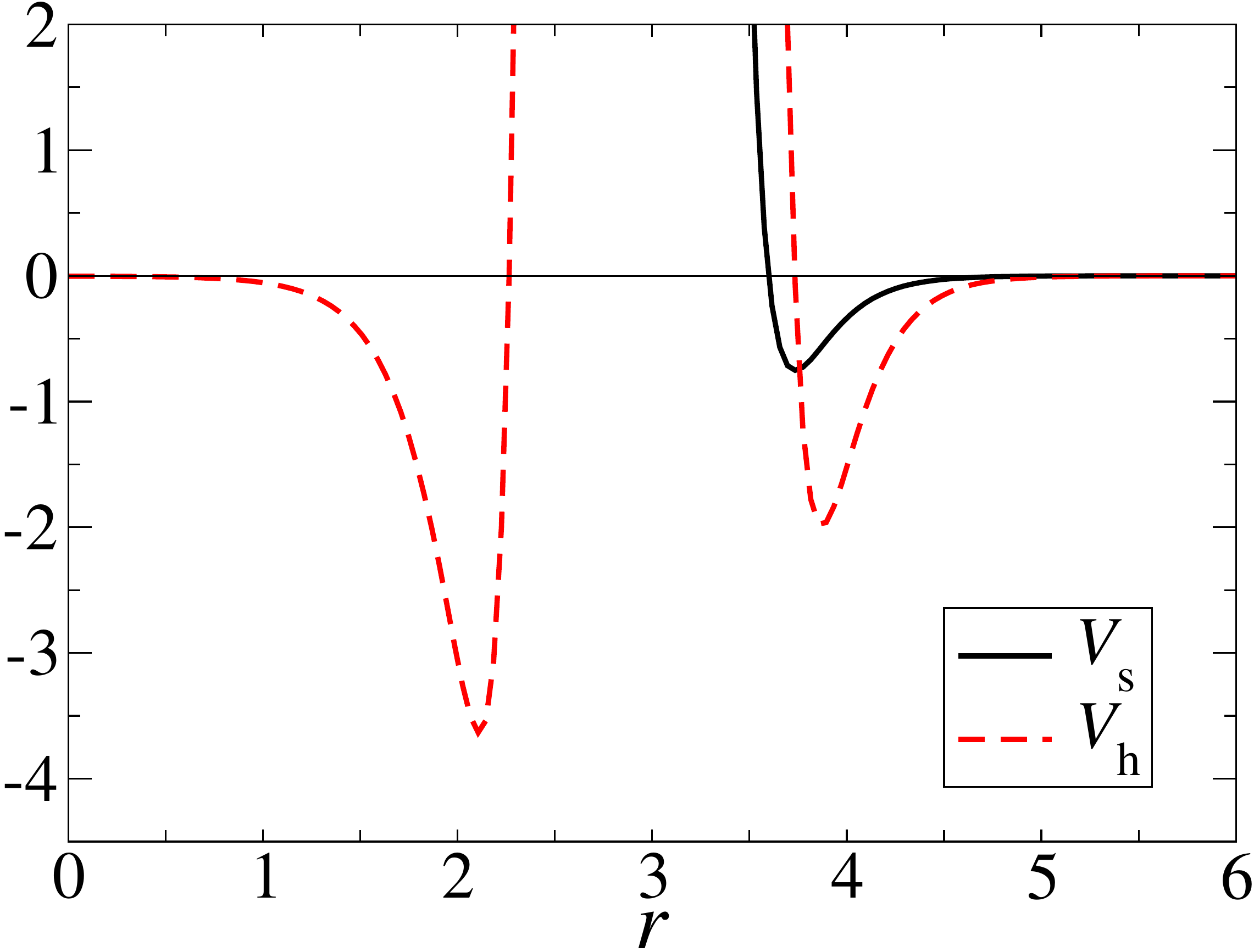}~ 
\includegraphics[width=.48\columnwidth]{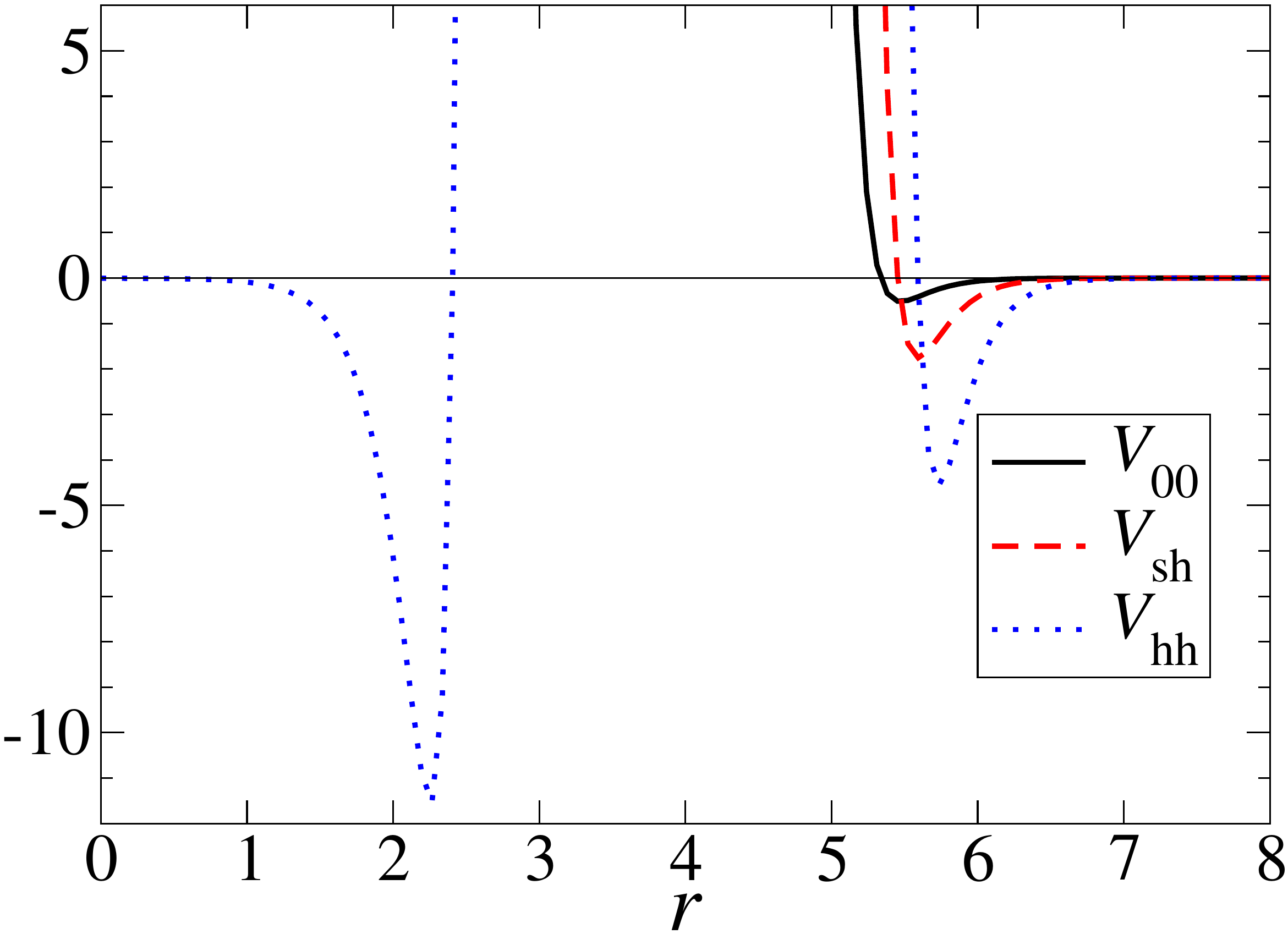}
\caption{Example of Morse effective potentials for $b=5.6$, for which
  the Morse potential resembles the Lennard-Jones potential. The left
  panel shows the effective point-nanoparticle potentials for
  $s=3$, the right panel shows the effective potentials for two
  nanoparticles of radius $s_1=4$ and $s_2=1$. Note that
  $V_\mathrm{sh}$ and $V_\mathrm{hs}$ are nearly the same for $r>D$.}
\label{Morseplot56}
\end{figure}

Two examples of the Morse-based effective potentials are shown in
Figs.~\ref{Morseplot26} and \ref{Morseplot56}, for $b=2.6$ and
$b=5.6$, respectively. For the lower value of $b$, there is a low
barrier for a point particle to penetrate a nanoparticle as well as
for one nanoparticle to penetrate another
(cf.~Fig.~\ref{Morseplot26}), while for the larger value of $b$ this
is virtually impossible (cf.~Fig.~\ref{Morseplot56}) if the energies
of the particles are of order~$1$.

\subsubsection*{Buckingham potential}

The modified Buckingham potential\cite{Buckingham38}
\begin{equation}
     \phi^\mathrm{B}(r) 
     = \left\{\begin{array}{ll}
          \infty               & \mbox{if $r<r^*$,}\\
           a e^{-br}-cr^{-6} & \mbox{if $r>r^*$,}
              \end{array}\right.
\label{Buckingham}
\end{equation}
is made up of an exponential part, for which the results of
Sec.~\ref{sec:expcluster} apply, and an attractive London-van der
Waals term treated above. In addition, one needs to take the cut-off
$r^*$ into account. This cut-off is necessary because otherwise, for
small enough $r$, the Buckingham potential would become
negative. Thus, the effective point-nanoparticle potentials are
\begin{eqnarray}
    V^\mathrm{B}_0(r) & = & \left\{\begin{array}{ll}
                      \infty & \mbox{if $r<s+r^*$}
		   \\
		       \frac{a}{b^3}V^\mathrm{E}_0(br,bs) -c V^{6}_0(r)
			     & \mbox{if $r>s+r^*$}
                   \end{array}\right.
\label{VBs}
\\
    V^\mathrm{B}_\mathrm{h}(r) & = & \left\{\begin{array}{ll}
			 \frac{a}{b^2}V^\mathrm{E}_\mathrm{h}(br,bs)-cV^{6}_\mathrm{h}(r)
			   & \mbox{if $r<s-r^*$}
	           \\
                         \infty & \mbox{if $|s-r|<r^*$}
		   \\
			 \frac{a}{b^2}V^\mathrm{E}_\mathrm{h}(br,bs) -c V^{6}_\mathrm{h}(r)
			 & \mbox{if $r>s+r^*$}
    \end{array}\right.
\label{VBh}
\end{eqnarray}
while the effective inter-nanoparticle potentials are given by
\begin{eqnarray}
    V_{00}^\mathrm{B}(r) & = & \left\{\begin{array}{ll}
                               \infty&\mbox{ if $r < D+r^*$}
			       \\
                               \frac{a}{b^6}V_{00}^\mathrm{E}(br,bs_1,bs_2)
                               -cV^{6}_{00} (r)
			       &\mbox{ otherwise}
			     \end{array}\right.
\label{VBss}
\\
    V_\mathrm{sh}^\mathrm{B}(r) & = & \left\{\begin{array}{ll}
                               \infty
                               &\mbox{ if $-d-r^* < r < D+r^*$}
			       \\
                               \frac{a}{b^5}V_\mathrm{sh}^\mathrm{E}(br,bs_1,bs_2)
                               -cV^{6}_\mathrm{sh} (r)
                               &\mbox{ otherwise}
    \end{array}\right.
\label{VBsh}
\\
    V_\mathrm{hh}^\mathrm{B}(r) & = & \left\{\begin{array}{ll}
                              \infty&\mbox{ if $|d|-r^*<r<D+r^*$}
			      \\
                              \frac{a}{b^4}V_\mathrm{hh}^\mathrm{E}(br,bs_1,bs_2)
                              -c V^{6}_\mathrm{hh}(r) 
		              &\mbox{ otherwise.}
    \end{array}\right.
\label{VBhh}
\end{eqnarray}
While the effective potentials due to the exponential pair potential
are different for different cases (no overlap, partial overlap, and
complete overlap), because of the presence of a cut-off $r^*$, only
the non-overlapping case is relevant here.

\begin{figure}[t]
\includegraphics[width=.48\columnwidth]{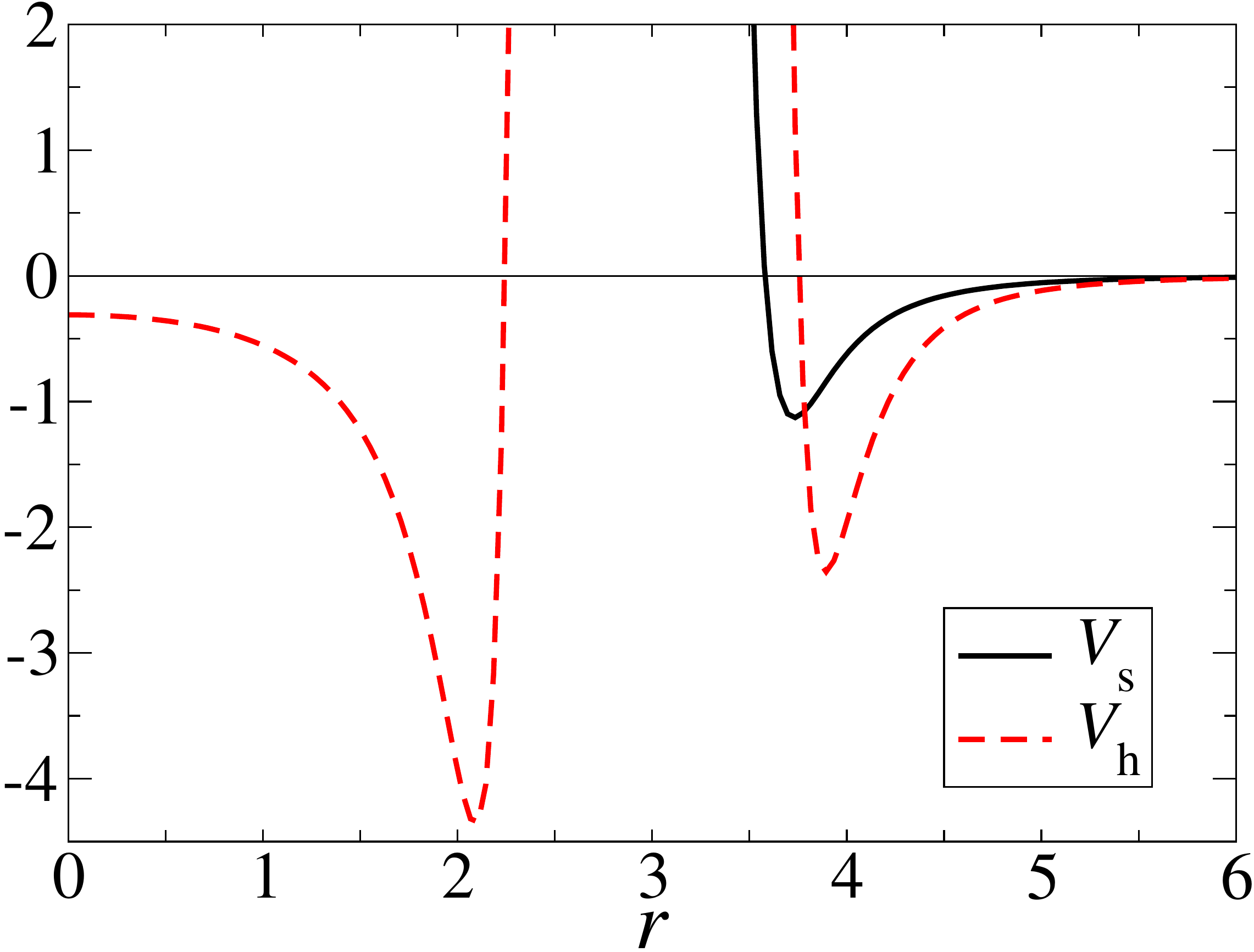}~ 
\includegraphics[width=.48\columnwidth]{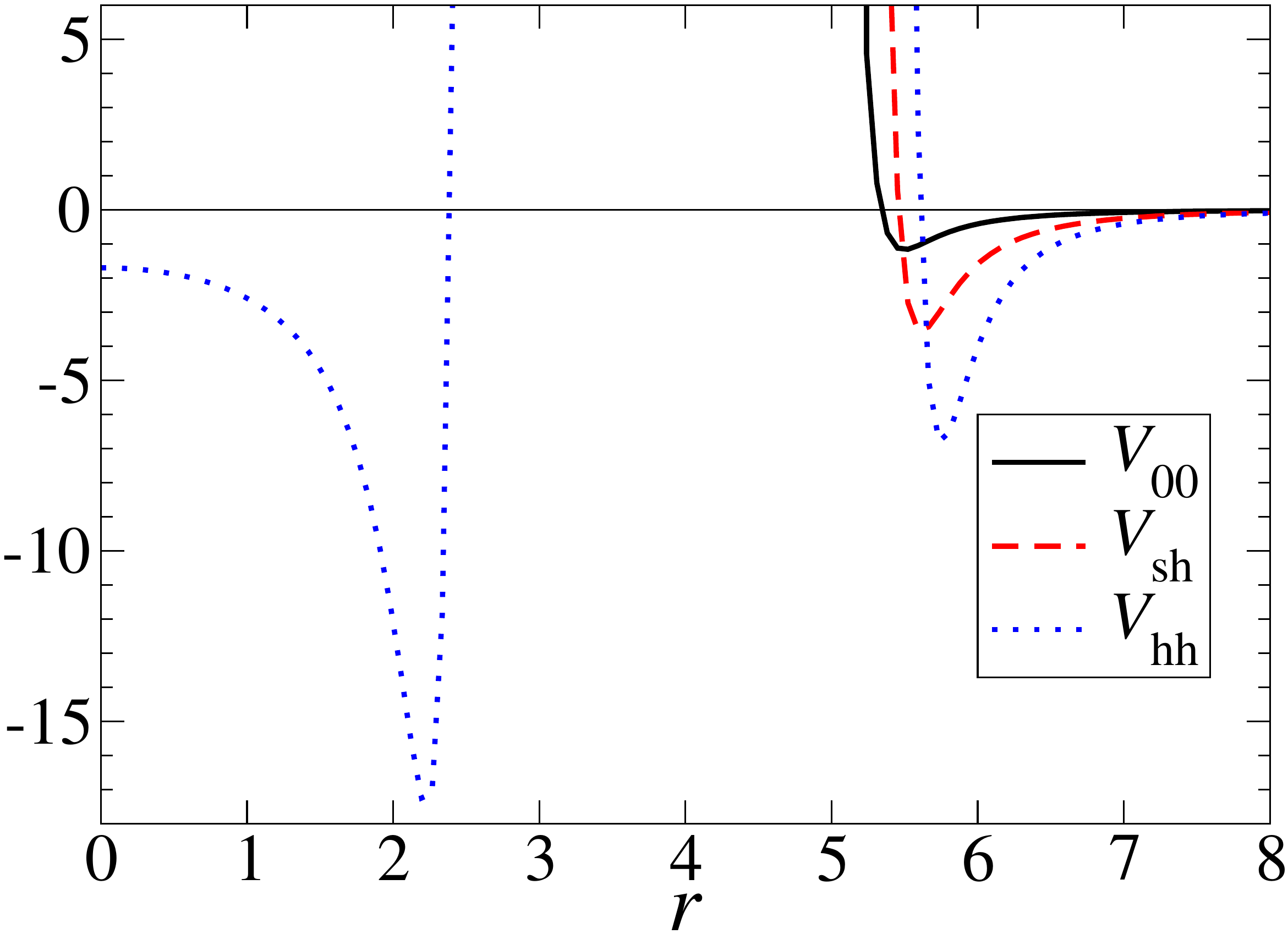}
\caption{Typical example of effective potentials based on the
  Buckingham potential for $a=e^{13}$, $b=13$ and $c=2$, with the
  cut-off $r^*$ set to $1/4$.  The left panel shows the effective
  potential for a point particle and a solid or hollow nanoparticle of
  radius $s=3$, the right panel shows the potentials for two
  nanoparticles of radius $s_1=4$ and $s_2=1$.}
\label{Bplot}
\end{figure}

In Fig.~\ref{Bplot}, a typical example of these potentials is
shown. Note that while it is possible for a point or nanoparticle
particle to be inside the hollow nanoparticle (as long as there is no
overlap), there is an infinite barrier to get inside from the outside,
in contrast with the effective potentials based on the Morse potential.

\subsubsection*{Lennard-Jones potential}

One of the most often used potentials in molecular dynamics
simulations is the Lennard-Jones potential,\cite{LennardJones37} which
in reduced units reads
\begin{equation}
     \phi^\mathrm{LJ}(r)
      =   \frac{1}{r^{12}}-\frac{2}{r^{6}}
  = \phi^{12}(r) -2  \phi^{6}(r).
\label{LJ}
\end{equation}
Since the attractive part of the Lennard-Jones potential in
Eq.~(\ref{LJ}) was handled above, one only needs to
add the repulsive part $r^{-12}$ to find the effective potentials for
Lennard-Jones nanoparticles. Substituting $n=12$ into the results of
Sec.~\ref{sec:powerlaws}, and using the relations between auxiliary
and effective potentials, one finds
\begin{eqnarray}
    V^{12}_0(r) &= &
        \frac{4\pi s^3}{3(r^2-s^2)^6}
        + \frac{80\pi s^9+432\pi r^4s^5}{45(r^2-s^2)^9}
\label{repV01}
\\
    V^{12}_\mathrm{h}(r) &= &
       \frac{4\pi s^2}{(r^2-s^2)^6}
       +\frac{64\pi r^2s^4(r^4+\frac65s^2r^2+s^4)}{(r^2-s^2)^{10}}
\label{repV02}
\\
    V^{12}_{00} (r)
    &=&  \frac{\pi^2}{37800r}
        \bigg[
           \frac{(r+\frac72D)^2+\frac54D^2-\frac{15}2d^2}{(r+D)^7}
           -\frac{(r+\frac72d)^2+\frac54d^2-\frac{15}2D^2}{(r+d)^7}
\nonumber\\&&\quad\qquad
          +\frac{(r-\frac72D)^2+\frac54D^2-\frac{15}2d^2}{(r-D)^7}
          -\frac{(r-\frac72d)^2+\frac54d^2-\frac{15}2D^2}{(r-d)^7}
        \bigg]
\label{repV11}
\\
    V^{12}_\mathrm{sh} (r)
    &=&  \frac{\pi^2s_2}{1260r}
        \bigg[
          -\frac{r+\frac92D+\frac72d}{(r+D)^8}
          -\frac{r-\frac92D-\frac72d}{(r-D)^8}
          +\frac{r+\frac92d+\frac72D}{(r+d)^8}
          +\frac{r-\frac92d-\frac72D}{(r-d)^8}
  \bigg]
\label{repV12}
\\
    V^{12}_\mathrm{hh}(r) & = &\frac{2\pi^2s_1s_2}{45r}
                      \bigg[
                         \frac{1}{(r+D)^9}+ \frac{1}{(r-D)^9}
                        -\frac{1}{(r+d)^9}- \frac{1}{(r-d)^9}
                      \bigg].
\end{eqnarray}
The potential $V_{00}^{12}$ is in agreement with the result in the
appendix of Ref.~\onlinecite{SchwarzSafran00}.

\begin{figure}[t]
\includegraphics[width=.48\columnwidth]{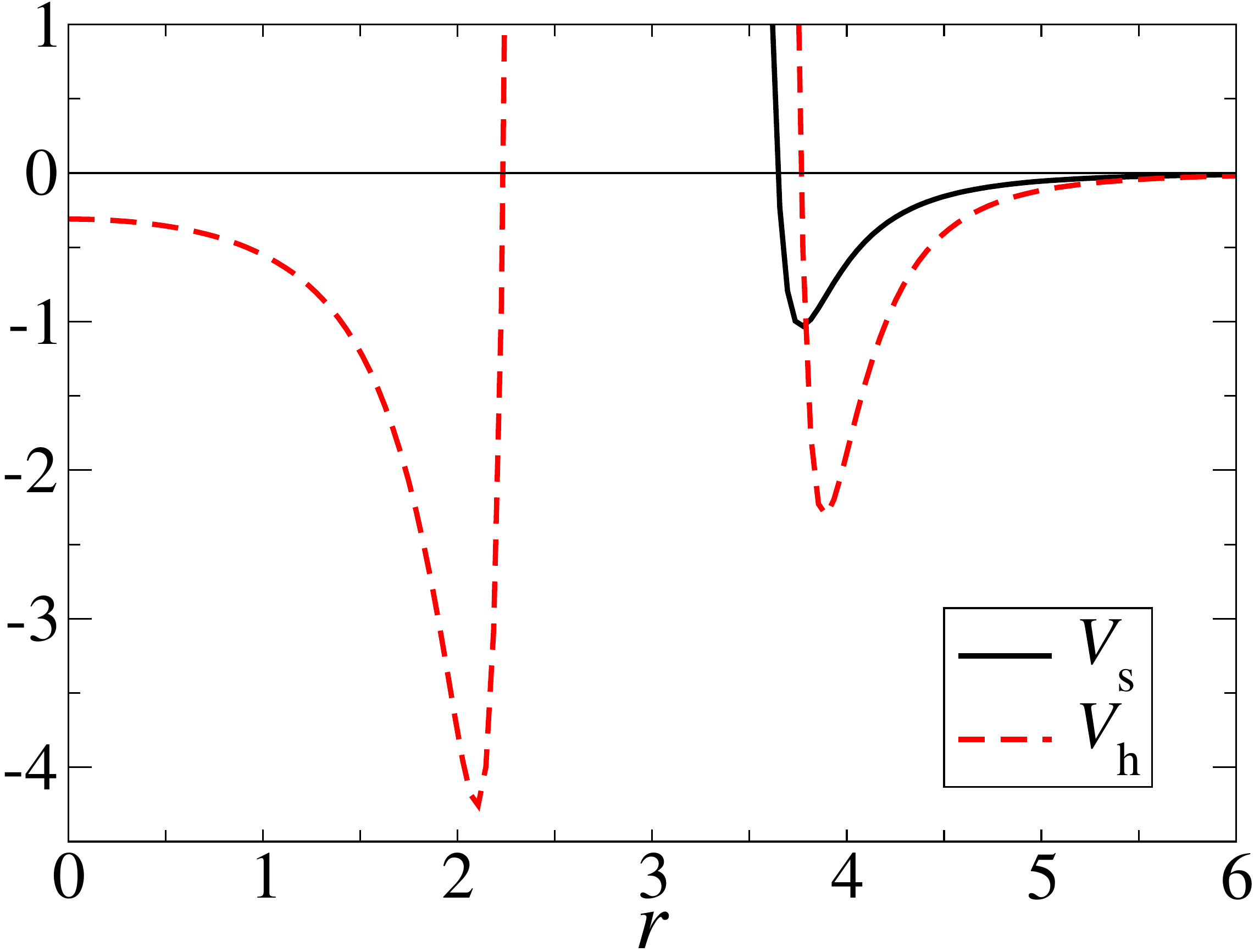}~ 
\includegraphics[width=.48\columnwidth]{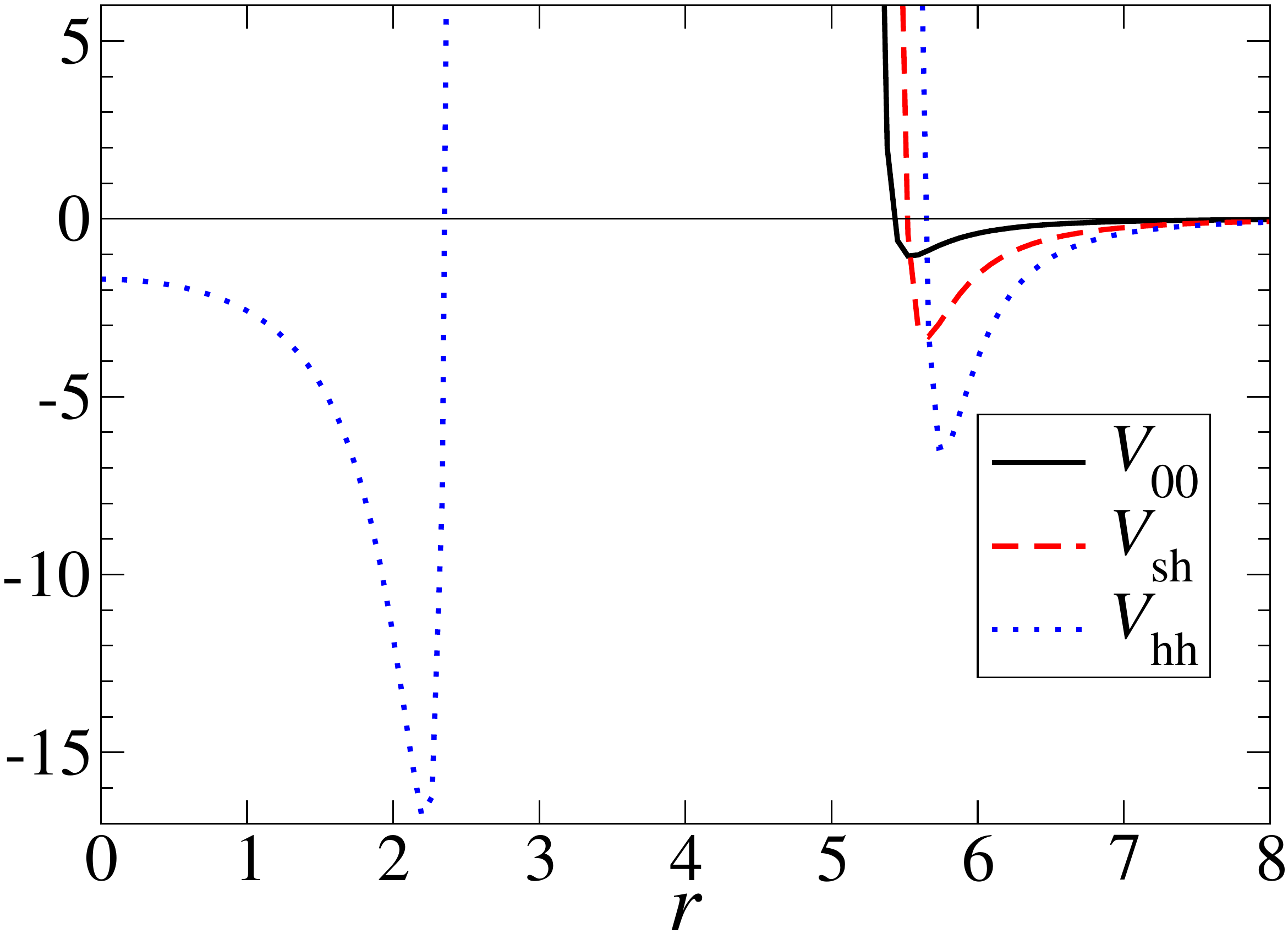}
\caption{Typical effective potentials based on the Lennard-Jones
  potential [Eqs.~(\ref{VLJsexplicit})--(\ref{VLJij})]. One the left, the
  potential for a point particle and solid or hollow nanoparticle of
  radius $s=3$ is shown, and on the right, the potentials for two
  nanoparticles of radius $s_1=4$ and $s_2=1$.}
\label{LJplot}
\end{figure}

The point-nanoparticle potentials for the Lennard-Jones potential are
now given by
\begin{eqnarray}
  V^\mathrm{LJ}_0(r) 
  & = &
  V_0^{12}(r) -2 V_0^{6}(r)
\nonumber\\
  & = &
  \frac{4\pi s^3}{3(r^2-s^2)^6}
  +\frac{80\pi s^9+432\pi r^4s^5}{45(r^2-s^2)^9}
  - \frac{8\pi s^3}{3(r^2-s^2)^3} 
\label{VLJsexplicit}
\\
  V^\mathrm{LJ}_\mathrm{h}(r) 
  & = &
  V_\mathrm{h}^{12}(r) -2 V_\mathrm{h}^{6}(r)
\nonumber\\
  &=&
  \frac{4\pi s^2}{(r^2-s^2)^6} +\frac{64\pi
  r^2s^4(r^4+\frac65s^2r^2+s^4)}{(r^2-s^2)^{10}}
                     -\frac{8\pi s^2}{(r^2-s^2)^3}
                    - \frac{16\pi s^4}{(r^2-s^2)^4}.
\label{VLJhexplicit}
\end{eqnarray}
Equation (\ref{VLJsexplicit}) is a more concise notation of the result
of Roth and Balasubramanya [Eq.~(2) in
Ref.~\onlinecite{RothBalasubramanya00}].  Likewise, the
inter-nanoparticle interactions due to a Lennard-Jones potential are
given by
\begin{equation}
    V_{ij}^\mathrm{LJ}(r) = V_{ij}^{12}(r) - 2  V_{ij}^{6}(r),
\label{VLJij}
\end{equation}
where $ij= 00$, sh or hh. In Fig.~\ref{LJplot}, a typical example of
these effective potentials is shown.  Note the hard core part of the
potentials. For the specific case of system of nanoparticles with the
same radii $s_1=s_2=s$, studied in
Ref.~\onlinecite{VanZonAshwinCohen08}, the effective
inter-nanoparticle interactions can be written in terms of $\eta=r/s$
as
\begin{eqnarray}
    V^\mathrm{LJ}_{00}(r) & = &
    \frac{\pi^2\sum_{i=0}^5\,\alpha_i^\mathrm{ss} \,\eta^{2i}}{s^6\,\eta^8\,(\eta^2-4)^7}
     -   \frac{4\pi^2}{3}
       \frac{\eta^2-2}{\eta^2\,(\eta^2-4)}
     -  \frac{\pi^2}{3}
       \ln\Big(1-\frac{4}{\eta^2}\Big)
\label{VLJssexplicit}
\\
   V^\mathrm{LJ}_\mathrm{sh}(r) & =&
    \frac{\pi^2\sum_{i=0}^6\,\alpha_i^\mathrm{sh}\, \eta^{2i}}{s^7\,\eta^8\,(\eta^2-4)^8}
    - \frac{32\pi^2}{3s\, \eta^2\,(\eta^2-4)^2}
\label{VLJshexplicit}
\\
   V^\mathrm{LJ}_\mathrm{hh}(r)  &=&
    \frac{\pi^2\sum_{i=0}^8\,\alpha_i^\mathrm{hh} \,\eta^{2i}}{s^8\,\eta^{10}\,(\eta^2-4)^9}
   -32\pi^2\frac{\eta^4+6\eta^2-8}{s^2\,\eta^4\,(\eta^2-4)^3}
\label{VLJhhaexplicit}
\end{eqnarray}
with the $\alpha$ coefficients given in Table~\ref{coefficients}. Equation
(\ref{VLJhhaexplicit}) is the so-called Girifalco
potential.\cite{Girifalco92}

\begin{table}[t]
\centering
\begin{tabular}{c|ccccccccc}
$i:$&0&1&2&3&4&5&6&7&8
\\\hline
$\alpha_i^\mathrm{ss}$&$-\frac{2^{13}}{315}$&$\frac{219136}{4725}$&$-\frac{24064}{675}$&$\frac{3456}{225}$&$-\frac{2^7}{45}$&$\frac{2^4}{9}$\\
$\alpha_i^\mathrm{sh}$&$\frac{2^{15}}{315}$&$-\frac{2^{16}}{315}$&$\frac{2^{13}}{45}$&$-\frac{2^{12}}{45}$&$\frac{2^7}{3}$&$-\frac{2^8}{15}$&$\frac{2^4}{3}$\\
$\alpha_i^\mathrm{hh}$&$\frac{2^{20}}{45}$&$-\frac{2^{18}}{5}$&$\frac{2^{18}}{5}$&$-\frac{917504}{30}$&$\frac{57344}{5}$&$-\frac{13312}{5}$&$\frac{14336}{15}$&$2^7$&$2^4$
\end{tabular}
\caption{Coefficients for the polynomials appearing in the effective
  inter-nanoparticle potentials based on the Lennard-Jones potentials,
  i.e., $V_{00}^\mathrm{LJ}$, $V_\mathrm{sh}^\mathrm{LJ}$ and
  $V_\mathrm{hh}^\mathrm{LJ}$ in
  Eqs.~(\ref{VLJssexplicit})--(\ref{VLJhhaexplicit}).
  \label{coefficients}}
\end{table}

\section{Accuracy of the Lennard-Jones based effective potentials for
  fcc nanoparticles} 
\label{sec:comparison}

\begin{figure}[t]
\includegraphics[width=.9\columnwidth]{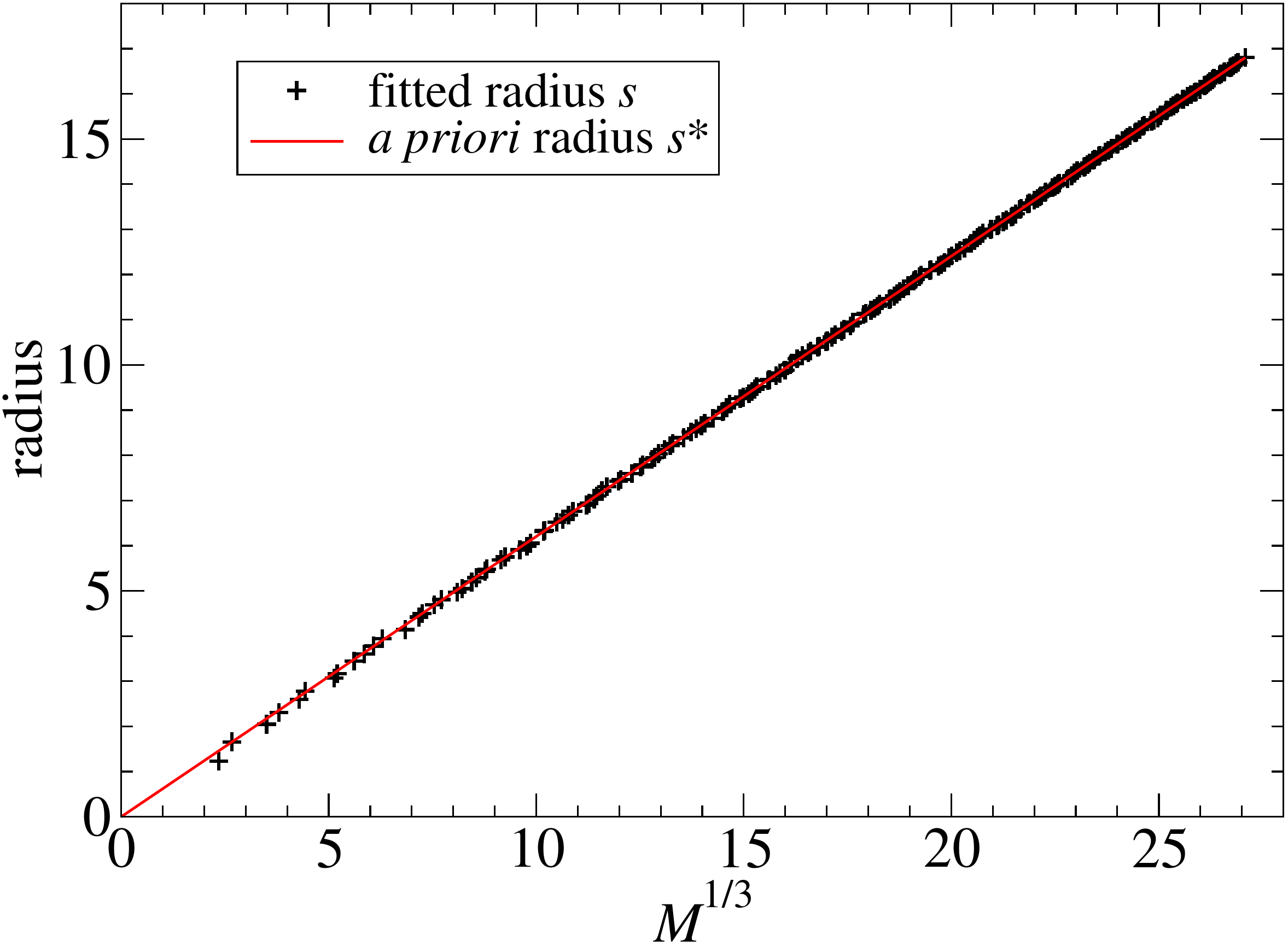}
\caption{Comparison of the fitted radius $s$ and the \textit{a
priori} radius $s^*$ of the fcc nanoparticles. The fit is based on
minimizing $\Delta_\mathrm{pn}(s)$, but minimizing
$\Delta_\mathrm{nn}(s)$ instead gives indistinguishable results.
\label{sfitvss}}
\end{figure}

Since the effective potentials derived above are intended to model
nanoparticles, it is natural to ask to what extent they can represent
the interactions of nanoclusters composed of atoms. This obviously
will depend on the structure of the nanoclusters, but to get at least
a partial answer, the fcc-based nanoparticles of
Sec.~\ref{sec:general} will be used again, with the basic pair
potentials $\phi_\mathrm{pn}$ and $\phi_\mathrm{nn}$ given by
$\phi^\mathrm{LJ}$ in Eq.~(\ref{LJ}). This potential has a minimum at
$r=1$, which sets the unit of length.  The fcc nanoparticles are
constructed from an fcc lattice with mean density $\bar\rho=1$ by
picking an atom and including all atoms within a given distance from
it.  Note that this gives only specific values for the number $M$ of
included atoms, since many atoms lie at the same distance in the
crystal structure.  Here, $M$ will be restricted to less than
20,000, resulting in 206 clusters, the largest of which has
$M=$19,861 atoms.

The mean density $\bar\rho=1$ for the fcc nanoparticles is not
unrealistic: It results in a lattice distance $a=4^{1/3}$
(Ref.~\onlinecite{Kittel86}, p.\ 12), i.e., the ratio of the lattice
distance to the interaction range is $4^{1/3}\approx 1.587$. This is
comparable to the case of platinum nanoparticles in water: Assuming the
lattice distance $a$ is the same as in a bulk platinum crystal,
$a=3.92$~\AA\ (Ref.~\onlinecite{Kittel86}, p.\ 23), and using that the
interaction range of Pt atoms with water is of the order of 2 to
3~\AA,\cite{Spohr89} one finds a similar ratio of
$3.92\mbox\AA/2.5\mbox\AA=1.568$.

\begin{figure}[t]
\includegraphics[width=.9\columnwidth]{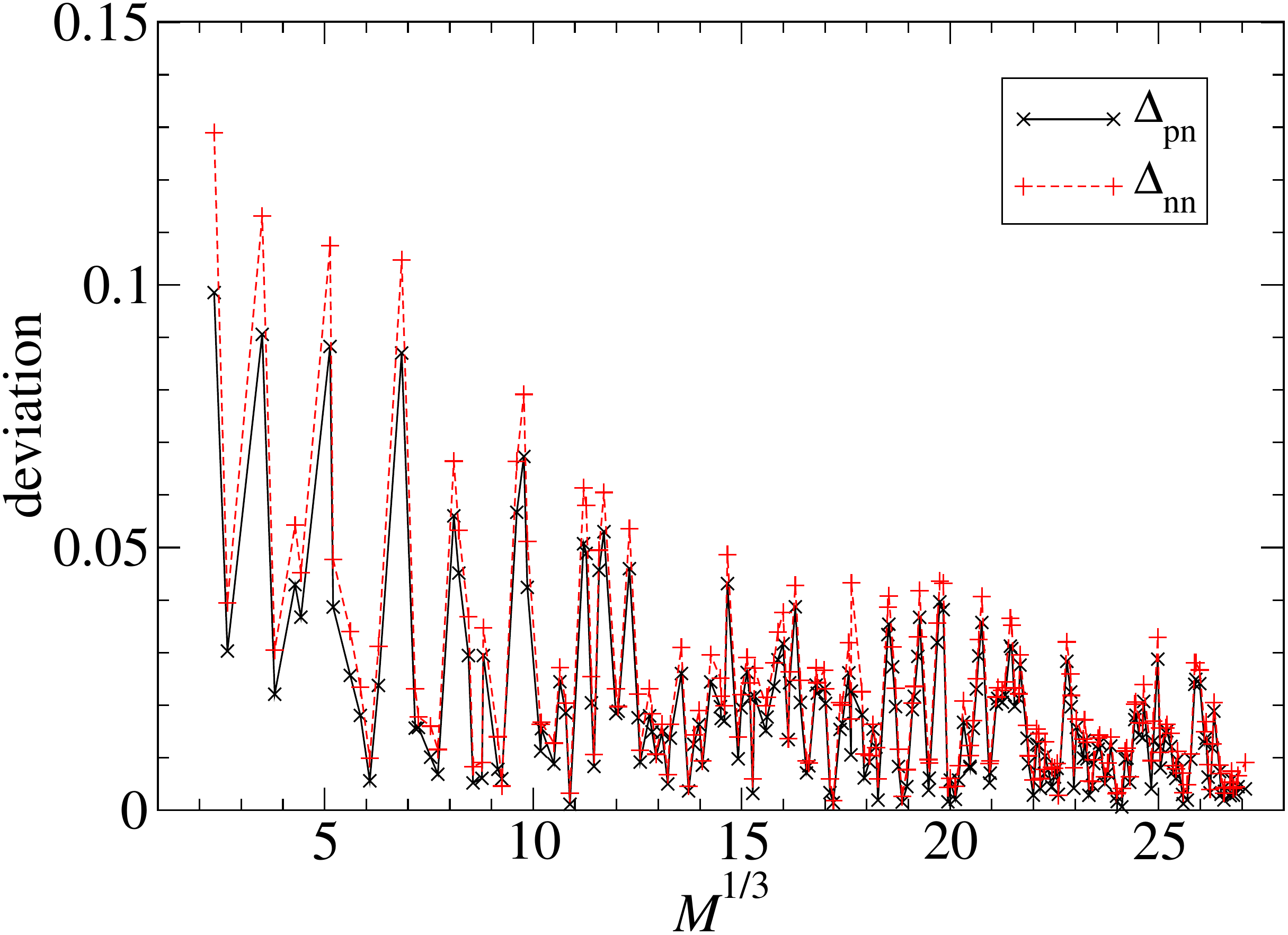}
\caption{Deviations $\Delta_\mathrm{pn}$ and $\Delta_\mathrm{nn}$ of
  the effective potentials from the atom-by-atom summed potentials for
  the fcc nanoparticles as a function of
  the fitted radius $s$.\label{delta}}
\end{figure}

To test the applicability of describing these fcc nanoclusters as
spheres with a constant density, one should compare the effective
point-nanoparticle potential $V_\mathrm{pn}=\rho V^\mathrm{LJ}_0$ to the
result of summing the potentials $\phi^\mathrm{LJ}$ between the point
particle and each of the atoms in the fcc nanoparticle. Similarly, the
effective potential $V_\mathrm{nn}=\rho^2V^\mathrm{LJ}_{00}$ between two
equally sized nanoparticles should be compared to the result of summing
the potentials between the each of the atoms of one of the
nanoparticles with each of the atoms in the other.  

However, there are two difficulties in performing these
comparisons. First, the effective potentials are spherically
symmetric, but the summed potentials will not be, since the fcc
nanoparticles are not truly spherically symmetric.  Therefore, the
comparison will be made with the summed potentials averaged over all
orientations of the nanoparticles, which will be denoted by
$V^\mathrm{sum}_\mathrm{pn}$ and $V^\mathrm{sum}_\mathrm{nn}$.

The second problem with the comparison is that the radius $s$ of the
nanoparticle, which is a parameter in the effective potentials, is not well
defined.  A reasonable \textit{a priori} radius would be $s^*=[3M/(4
\pi\bar\rho)]^{1/3}$, but other values for the radius $s$ close to
$s^*$ are just as reasonable.  Thus, the radius may be viewed as a
fitting parameter, which will be adjusted to minimize the difference
between the effective and the summed potential. To be precise, the
following quantities are minimized by varying $s$:
\begin{eqnarray}
  \tilde\Delta_\mathrm{pn} &=& \left\{\int'\!\mathrm dr\:
        \left[V^\mathrm{sum}_\mathrm{pn}(r)-\rho(s) V^\mathrm{LJ}_0(r)\right]^2\right\}^{1/2}
\nonumber\\
  \tilde\Delta_\mathrm{nn} &=& \left\{\int'\!\mathrm dr\:
        \left[V^\mathrm{sum}_\mathrm{nn}(r)-\rho^2(s)V^\mathrm{LJ}_{00}(r)\right]^2\right\}^{1/2}
\label{deltadef}
\end{eqnarray}
Here, $\rho(s)=3M/(4\pi s^3)$, and the prime denotes the restriction
on the integration that $V_\mathrm{pn}^\mathrm{sum}(r) < 3
V^*_\mathrm{pn}$ or $V_\mathrm{nn}^\mathrm{sum}(r) < 3
V^*_\mathrm{nn}$, respectively, where $V^*_\mathrm{pn}$ and
$V^*_\mathrm{nn}$ are the absolute value of the minima of
$V_\mathrm{pn}^\mathrm{sum}$ and $V_\mathrm{nn}^\mathrm{sum}$. The
restriction is needed to make the integrals converge.  The results
depend very little on the precise choice of the restriction. For
instance, changing the restriction to $2V^*$ instead of $3V^*$, shifts
the values for the radii $s$ only by an amount of the order of
$10^{-4}$.

The values of the radius that result from minimizing
$\Delta_\mathrm{pn}$ for the 206 cluster configurations with $M<$20,000
are shown in Fig.~\ref{sfitvss}. It is seen that with the exception of
some of the smaller clusters, the values of fitted radii $s$ typically
lie close very to the \textit{a priori} radius $s^*$.  Minimizing
$\Delta_\mathrm{nn}$ instead results in the same values for the radii
to within $0.3\%$.

\begin{figure}[t]
\includegraphics[width=.48\columnwidth]{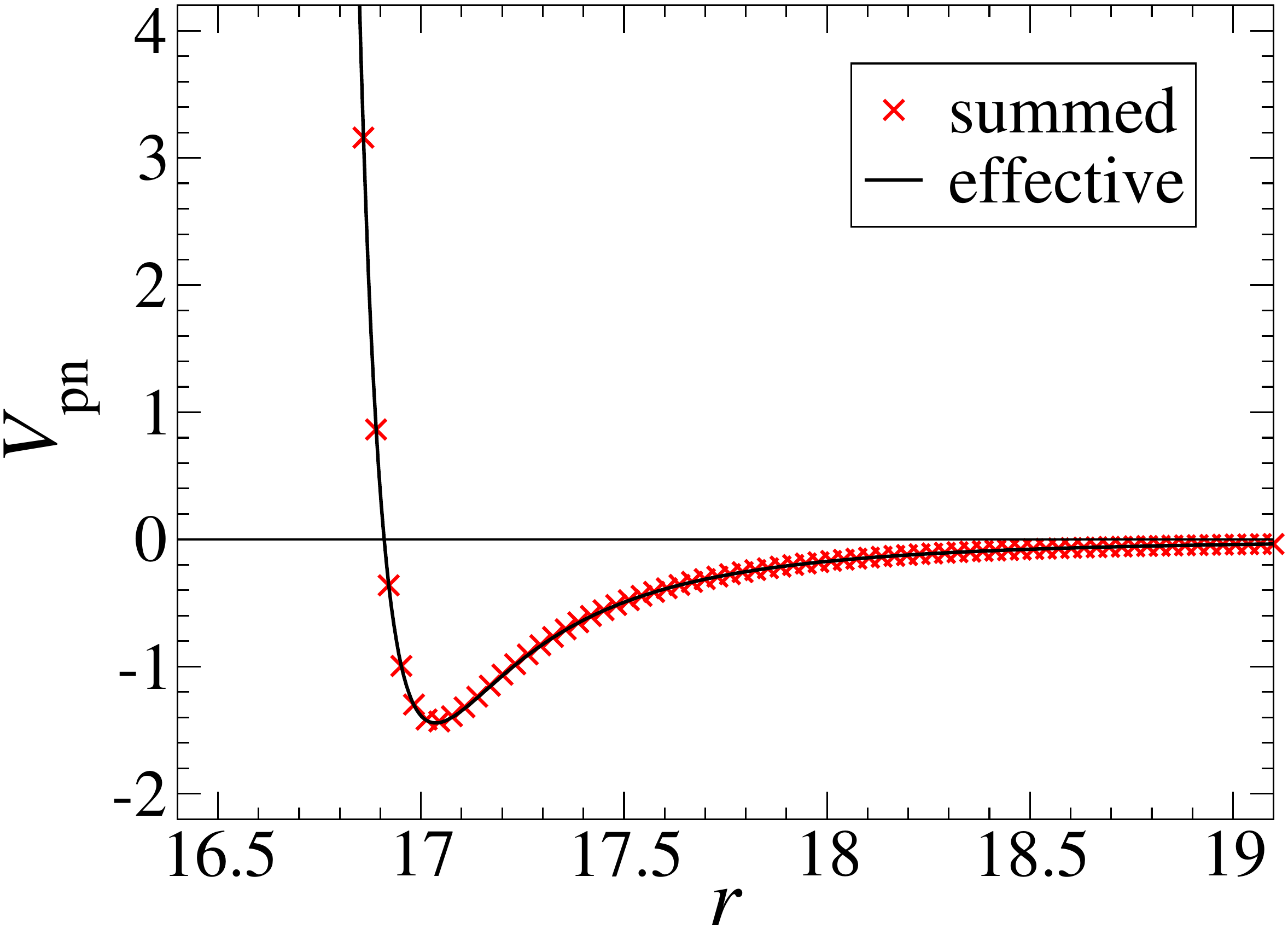}~ 
\includegraphics[width=.48\columnwidth]{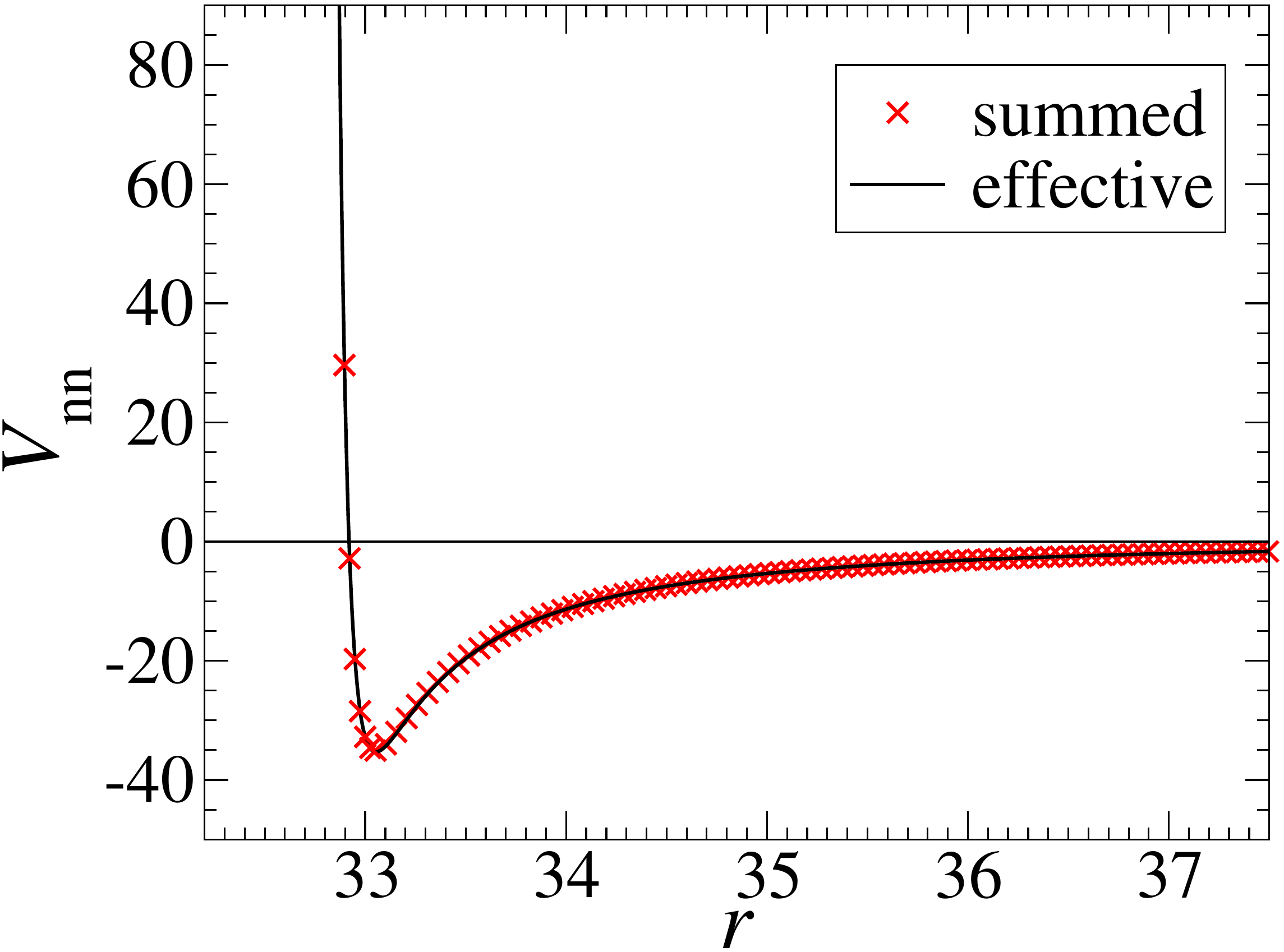}
\caption{An example of very good agreement between the effective and
  summed potentials, which occurs for an fcc nanocluster of size
  $M=18053$ with an effective radius of $16.27$ (in dimensionless
  units). Crosses represent the orientationally averaged summed
  potentials $V^\mathrm{sum}_\mathrm{pn}$ (left) and
  $V^\mathrm{sum}_\mathrm{nn}$ (right), while the solid lines are the
  effective potentials $V_\mathrm{pn}=\rho V^\mathrm{LJ}_{0}$ (left)
  and $V_\mathrm{nn}=\rho^2 V^\mathrm{LJ}_{00}$ (right). 
\label{examples1}}
\end{figure}

\begin{figure}[t]
\includegraphics[width=.48\columnwidth]{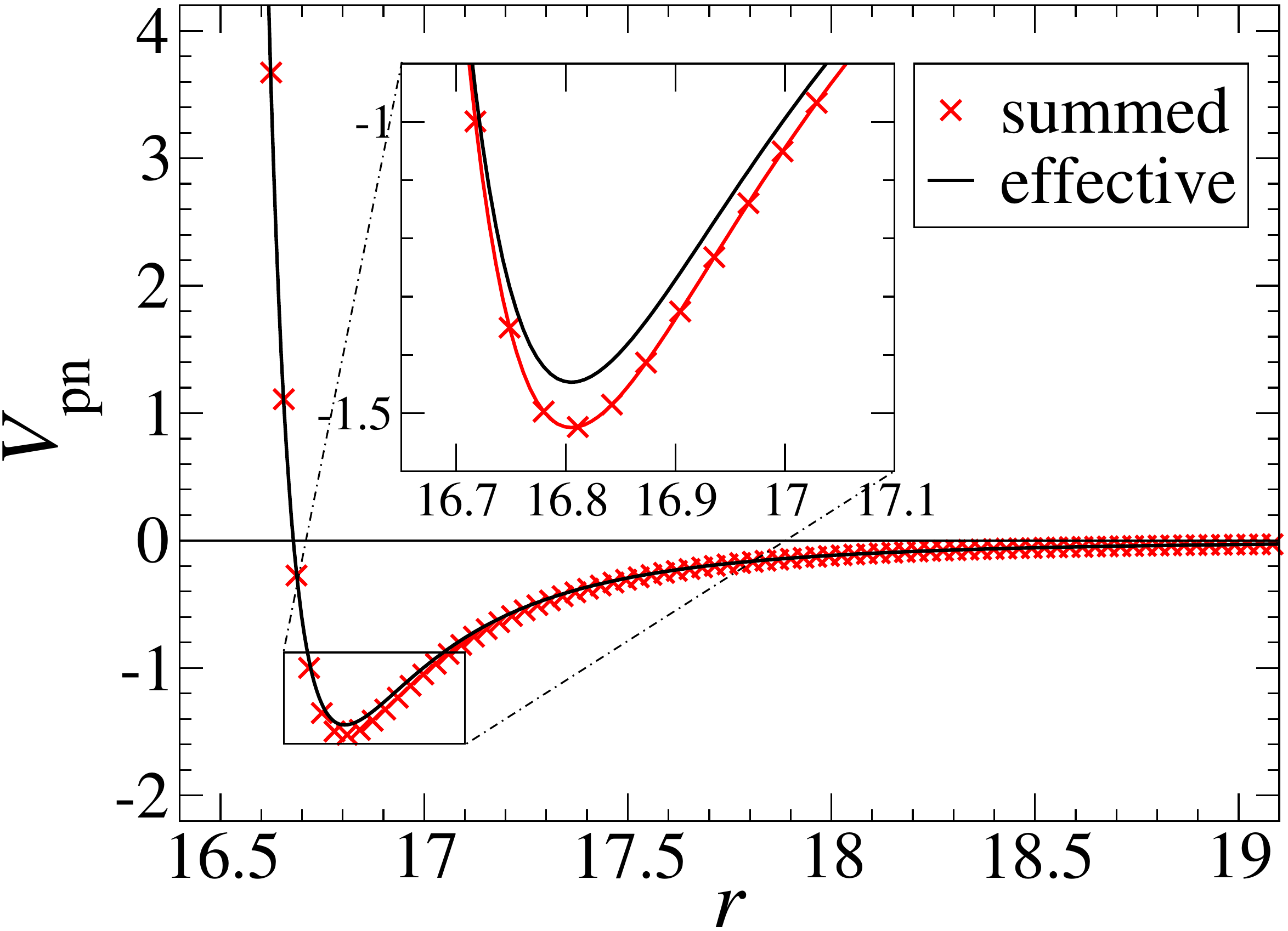}~ 
\includegraphics[width=.48\columnwidth]{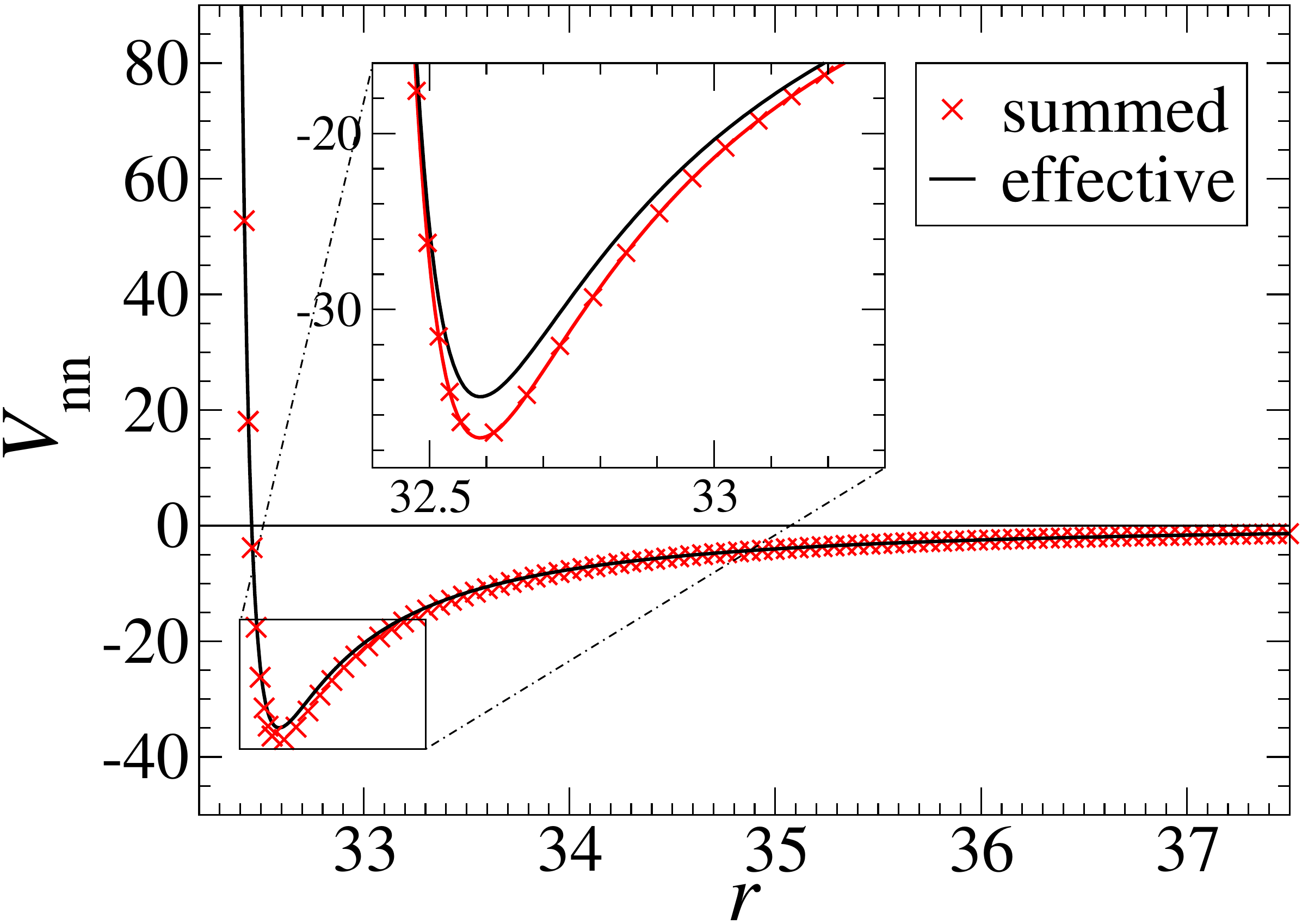}
\caption{An example of poorer agreement between the effective and
  summed potentials, which occurs for an fcc nanocluster of size
  $M=17357$ with an effective radius of $16.04$ (dimensionless
  units). Solid lines represent the effective potentials
  $V_\mathrm{pn}=\rho V^\mathrm{LJ}_{0}$ (left) and
  $V_\mathrm{nn}=\rho^2 V^\mathrm{LJ}_{00}$ (right), while crosses are
  the orientationally averaged summed potentials
  $V^\mathrm{sum}_\mathrm{pn}$ (left) and $V^\mathrm{sum}_\mathrm{nn}$
  (right) which result from the sum over the atoms.  The inset in the
  right plot zooms in on the minimum, and shows that its depth is
  underestimated by the effective potential.
\label{examples2}}
\end{figure}

To get an idea of the accuracy of the fit as a function of the size of
the nanoparticles, one may investigate the values of the dimensionless
deviations
\begin{eqnarray*}
  \Delta_\mathrm{pn} =
  \frac{\tilde\Delta_\mathrm{pn}}{R^{1/2}_\mathrm{pn}V^*_\mathrm{pn}};
&&
  \Delta_\mathrm{nn} =
  \frac{\tilde\Delta_\mathrm{nn}}{R^{1/2}_\mathrm{nn}V^*_\mathrm{nn}}.
\end{eqnarray*}
The length scales $R_\mathrm{pn}$ and $R_\mathrm{nn}$ are chosen as
the lengths of the intervals contributing $99.9\%$ of the values of
the integrals in Eqs.~(\ref{deltadef}). This typically gives
$R_\mathrm{pn}\approx1.35$ and $R_\mathrm{nn}\approx2$ for the size of
clusters investigated here, and these values of $R_\mathrm{pn}$ and
$R_\mathrm{nn}$ were used for all clusters.  The dimensionless
deviations are plotted in Fig.~\ref{delta}. One sees a high degree of
correlation between the accuracy of the effective potential for a
nanoparticle and point particle and the accuracy of the effective
inter-nanoparticle potential.  The deviations are furthermore
typically small, indicating that their is good agreement between the
effective potentials and the sum of atom-atom potentials, although the
deviations are larger for specific cluster sizes.  As extreme
examples, Fig.~\ref{examples1} shows a case of very good agreement and
Fig.~\ref{examples2} shows a case of poor agreement. In these figures,
the effective potentials and the summed potentials are compared for
$M=18053$ with $s=16.27$ and $M=17357$ with $s=16.04$,
respectively. Note that the agreement is never very bad, but for the
latter, the depth of the minimum is somewhat underestimated by the
effective potentials, as the insets of Fig.~\ref{examples2} show.

It is hard to say in general why the smooth, constant density
description works better for some clusters than for others. For some
of the smaller nanoclusters with poorer agreement, inspecting the
spatial structure of the nanocluster shows a rather rough surface,
which could be the explanation. But for the larger nanoparticles, such
a difference in roughness is hard to distinguish.

\section{Discussion}
\label{sec:discussion}

A general effective description for nanoparticles was presented,
starting from a smoothing procedure in which the real spatial density
profile inside the nanoparticles is replaced by a spherically
symmetric one.  The resulting effective interactions between a
nanoparticle and a point particle as well as between two nanoparticles
are then given by spherically symmetric potentials, thus greatly
simplifying the description over an all-atom model.

The main results of this approach are the formulation of the effective
potentials in terms of auxiliary potentials, Eqs.~(\ref{auxiliaryi})
and (\ref{auxiliaryij}), which provide a unified description of
overlapping and non-overlapping configurations.  The auxiliary
potentials are related to the basic interaction potentials through
Eqs.~(\ref{Aidef}) and (\ref{Aijdef}).  Furthermore, the effective
potentials for hollow particles were found to be related to those for
solid nanoparticles by simple differentiation with respect to the
radii of the nanoparticles, see Eqs.~(\ref{Vhdef}) and
(\ref{Urelations}), and as such also allow a formulation in terms of
auxiliary potentials, as given in Sec.~\ref{sec:solid-hollow}.

As an application of the formalism, explicit effective pair potentials
for solid and hollow nanoparticles were obtained for various basic
pair potentials.  Different pair potentials have different
applications. For instance, the Lennard-Jones potential is a
general-purpose potential, while the Buckingham potential is suited to
describe the physics of particles close together such as in high
pressure systems. These basic potentials result in effective
nanoparticle potentials with hard cores plus a soft potential.  They
reduce in limiting cases to some of the existing model potentials for
colloids, such as hard spheres and the Hamaker
potential,\cite{Hamaker37,
DerjaguinLandau41VerweyOverbeek48CichockiFelderhof88CichockiHinsen90,
BarratHansen} but not to more ad hoc models such as the description of
a colloid as a single big Lennard-Jones particle.\cite{LeeKapral04} In
contrast, the Morse potential is able to describe bounded systems or
penetrable particles, making it possible to model nanoparticles that
could passively capture and trap specific types of particles. This
could have applications in modeling drug delivery by
nanoparticles\cite{Jurgonsetal06Arrueboetal06} and viral
capsids.\cite{capsids}

For the case of a Lennard-Jones basic potential, a comparison was
carried out with an atomic model of a nanocluster. In this model, the
atoms making up the nanoparticle were assumed to be arranged in an fcc
lattice structure.  To find approximate spherical structures, the
atoms were restricted to lie within a certain distance from the
central atom in the nanocluster.  Configurations with up to 19,861
atoms were studied. The effective potentials were compared with the
orientionally averaged sum of Lennard-Jones potentials due to the
individual atoms. The agreement tends to be very good, provided the
radius in the effective description is treated as a fitting
parameter. For some configurations, however, the fitting procedure
underestimates the depth of the minimum of the potentials. This may be
due to surface roughness of these structures, which is caused by
the imposed fcc structure and unlikely to be relevant for real
nanoclusters.

The application of the explicit expressions for the effective
potentials to numerical studies of spherical nanoparticles is in
principle straightforward.  In fact, the potentials in
Eqs.~(\ref{VLJsexplicit}) and (\ref{VLJssexplicit}) have already been
used in a numerical study of single particle transport in an
equilibrium nanofluid composed of solid nanoparticles and fluid
particles interaction through Lennard-Jones interactions, where the
validity of a Gaussian approximation of the Van Hove self-correlation
function was investigated, and found to hold up to picosecond time
scales for the fluid particles, and up to five to ten times longer
(depending on temperature) for nanoparticles with a size of
about~2~nm.\cite{VanZonAshwinCohen08}

Given the explicit expressions for the effective potentials, the
description allows a fairly direct route toward a qualitative model
for a given system of nanoparticles in a fluid, since reasonable
values for the parameters for commonly used pair potentials are
available in the literature,\cite{charmm} while the number of atoms in
a nanoparticle and its radius could be taken from experiments or
theoretical calculations.\cite{BalettoFerrando05} Furthermore, the
effective potentials have a physical range based on the interaction of
their constituents rather than on their radius.  Therefore, the
effective potentials that were derived here are expected to be useful
for the qualitative description of a wide variety of systems, from
mono-disperse nanoparticles in a fluid to mixtures of different kinds
of fluid particles, nanoclusters or buckyballs.

A number of interesting extensions present themselves for future
research. For instance, while the nanoparticles were assumed to be
composed of one kind of particle only, potentials for nanoparticles
composed of several types of particles can also be derived within the
current context if the distribution of the types is either
homogeneously mixed or distributed in spherical shells (so-called
core-shell nanoparticles\cite{Glotzeretal04,Molineroetal00}). The
spherical symmetry of the effective potentials, which decouples the
rotational and translational degrees of freedom, could be lifted to
extend the model to include rotational motion. This may be done by
adding interaction sites on the surface of the nanoparticle or a
multipole expansion.  As long as the orientationally dependent
potential is available, there are no obstacles in molecular dynamics
simulations of such
systems.\cite{VanZonSchofield07bVanZonOmelyanSchofield08} Furthermore,
combining the current model with the mesoscopic fluid model of
Malevanets and Kapral\cite{MalevanetsKapral99MalevanetsKapral00} would
yield a numerically efficient model of larger nanoparticles and
colloids that includes hydrodynamic effects. These avenues are
currently being investigated.

\acknowledgments

The author wishes to thank Profs.\ E.\ G.\ D.\ Cohen, R.\ Kapral, and
J.\ Schofield for useful discussions. This work was supported by the
National Sciences and Engineering Research Council of Canada 
and a Petroleum Research Fund from the American Chemical Society.

\appendix

\section{The kernel $K_{ij}$}

The integral in the expression for the kernel $K_{ij}$ in
Eq.~(\ref{Kijint0}) will be worked out now. Using Eq.~(\ref{kidef})
and the binomial formula for $(x-y)^{i+2}$, one finds, after
resummation, that
\begin{widetext}
\begin{eqnarray}
  K_{ij}(x,s_1,s_2)
  &=& 
  \int_{y_1}^{y_2}\!{\rm d}y\:
  \frac{4\pi^2\Theta(D-|x|)}{(i+2)(j+2)}\,
       [s_1^{i+2}-(x-y)^{i+2}][s_2^{j+2}-y^{j+2}]
\label{Kijint}
\\
  &=&
  \frac{4\pi^2\Theta(D-|x|)s_1^{i+2}s_2^{j+2}}{(i+2)(j+2)}
  \Bigg[y-\frac{s_1}{i+3}\left(\frac{x+y}{s_1}\right)^{i+3}
\nonumber\\&&
    +\frac{s_2}{j+3}\left(\frac{y}{s_2}\right)^{j+3}\left\{
    \left(\frac{x}{s_1}\right)^{i+2}
    F\left(-i-2,j+3;j+4;-\frac yx\right)-1\right\}
    \Bigg]_{y_1}^{y_2},
\label{Kijint2}
\end{eqnarray}
\end{widetext}
where $y_1=\max(-s_2,x-s_1)$ and $y_2=\min(s_2,x+s_1)$,
which are due to the finite support of the kernels $K_i$ and $K_j$,
and $F$ is the hypergeometric function.\cite{GradshteynRyzhik} Despite
its complicated appearance, Eq.~(\ref{Kijint2}) is simply a piecewise
polynomial in $x$ of degree $i+j+5$ at most. To see this, it is useful
to distinguish the following four non-trivial cases: \emph{case 1:}
$x>0$ and $|d|<|x|<D$, for which $y_1=x-s_1$ and $y_2=s_2$;
\emph{case 2:} $d>0$ and $|x|<|d|$, giving $y_1=-s_2$ and
$y_2=s_2$; \emph{case 3:} $d<0$ and $|x|<|d|$, giving
$y_1=x-s_1$ and $y_2=x+s_1$; and \emph{case 4:} $x<0$ and
$|d|<|x|<D$, for which $y_1=-s_2$ and $y_2=x+s_1$.  There are in
fact only two independent cases, because case \emph3 can be obtained
from the result of case \emph2 by interchanging $s_1$ and $s_2$
as well as $i$ and $j$ (which will also flip the sign of $d$), while
the result for case \emph4 can be obtained from that of case \emph1 by
setting $s_1$ to $-s_2$, $s_2$ to $-s_1$ and introducing a
minus sign, as can be proved by changing the integration variable in
Eq.~(\ref{Kijint}) from $y$ to $x-y$. Thus, one only needs to consider
the cases \emph1 and \emph2. Changing the integration variable from
$y$ to $z=s_2-y$ and using the binomial formula, 
Eq.~(\ref{Kijint}) for case \emph1 yields
\begin{eqnarray}
    K_{ij}(x,s_1,s_2) 
=
 (i+2)!(j+2)!
    \sum_{m=0}^{i+j+2} \frac{(D-x)^{m+3}}{(m+3)!} 
    \sum_{k=\max(1,m-j)}^{\min(i+2,m+1)}
    \frac{(-s_1)^{i+2-k}}{(i+2-k)!} 
    \frac{s_2^{j-m+k}}{(j-m+k)!} 
,
\label{Kij2}
\end{eqnarray}
which is polynomial in $x$ of degree $i+j+5$, while for case \emph2
the integral in Eq.~(\ref{Kijint}) can be found by using the binomial
formula for $(x-y)^{i+2}$, giving a polynomial of degree $i+2$, i.e.
\begin{eqnarray}
  K_{ij}(x,s_1,s_2) &=& \frac{8\pi^2s_2^{j+3}}{i+2}
\Bigg[\frac{s_1^{i+2}}{j+3}
-\sum_{k=0}^{i/2+1}
  {i+2 \choose 2k}\frac{s^{i+2-2k}_2x^{2k}}{(i+3-2k)(i+j+5-2k)} 
\Bigg].
\label{Kij3}
\end{eqnarray}

\end{document}